# PHOTOMETRIC OBSERVATIONS CONSTRAINING THE SIZE, SHAPE, AND ALBEDO OF 2003 EL61, A RAPIDLY ROTATING, PLUTO-SIZED OBJECT IN THE KUIPER BELT

short title: Photometric Observations of 2003 EL61

David L. Rabinowitz[1], Kristina Barkume[2], Michael E. Brown[2], Henry Roe[2], Michael Schwartz[3], Suzanne Tourtellotte[4], Chad Trujillo[5]

[1]Yale University, Center for Astronomy & Astrophysics, P.O. Box 208121, New Haven CT 06520-8121

[2]Division of Geological and Planetary Sciences, California Institute of Technology, Pasadena, CA 91125

[3]Tenagra Observatory, C2 Box 292, Nogales, Arizona, USA 85621

[4]Yale University, Astronomy Department, P.O. Box 208121, New Haven CT 06520-8121

[5]Gemini Observatory, 670 North A'ohoku Place, Hilo, HI 96720

ABSTRACT

We present measurements at optical wavelengths of the spectral reflectance, rotational light curve, and solar phase curve of 2003 EL61. With apparent visual magnitude 17.5 at 51 AU from the sun, this newly discovered member of the classical Kuiper Belt is now the third brightest KBO after Pluto and 2005 FY9. Our observations reveal an unambiguous, double-peaked rotational light curve with period 3.9154 ± 0.0002 hours and peak-to-peak amplitude 0.28 ± 0.04 mag. This is the fastest rotation period reliably determined for any body in the solar system larger than 100 km. Assuming the body has relaxed over time to the shape taken by a homogenous fluid body, our observations tightly constrain the shape and density. Given the mass we recently determined for 2003 EL61 from the orbit of a small satellite, we also constrain the size and albedo. We find a total length of 1960 to 2500 km, a mean density of 2600 to 3340 kg m$^{-3}$, and a visual albedo greater than 0.6. We also measure a neutral reflectance at visible wavelengths and a linear phase curve with slope varying from 0.09 mag deg$^{-1}$ in the B band to 0.13 mag deg$^{-1}$ in the I band. The absolute V-band magnitude is 0.444 ± 0.021.

Subject headings: Kuiper Belt --- minor planets, asteroids --- solar system: general

1. INTRODUCTION

   In December of 2004, we discovered the bright Kuiper Belt object 2003 EL61 during the course of an on-going survey for distant solar-system objects with the 160-megapixel Quest camera on the 48" Samuel Oschin Schmidt telescope at Palomar (Brown, Trujillo & Rabinowitz 2004). At a distance of 51 AU and with mean V magnitude of 17.5, 2003 EL61 is currently the third brightest trans-Neptunian object after Pluto and 2005 FY9 and adds to a growing list of Pluto-scale KBOs. It is a member of the classical Kuiper Belt, with semi-major axis 43.3 AU, eccentricity 0.19, and inclination 28.2 deg. Our recent observations of 2003 EL61 using Keck adaptive optics with laser guide star reveal a much fainter companion satellite with near circular orbit and 49-day period (Brown et al 2005). From these observations, we determine a mass m = 4.21±0.1 x 10$^{21}$ kg for 2003 EL61, which is 32% the mass of Pluto (the mass of the satellite is negligible). We have also measured near-infrared reflectance spectra with the Gemini 8-m and Keck 10-m telescopes that show a clear signature of water ice (Trujillo et al. 2005).

   In this paper we present BVRI measurements of the color and of the solar phase curve for 2003 EL61. Combined with the infrared reflectance spectra, these results help to characterize the composition and structure of the surface. For example, our recent observations of 2003 UB313 (Brown, Trujillo, and Rabinowitz 2005) reveal a surface composition very much like Pluto's, dominated by methane ice, but lacking Pluto's red color at visible wavelengths. The absence of the red color may be an indication that methane ice is more evenly distributed over the surface of 2003 UB313 than it is on Pluto, masking underlying areas of red color. With the addition of 2003 EL61, we now have a handful of 1000-km-sized bodies whose photometric properties we can compare to each other and to smaller TNOs. This will help to address issues related to icy surfaces, the source of the red color on Pluto and smaller TNOs, and the range of albedos observed for the smaller population.

In this paper we also determine the period and amplitude of the rotational light curve for 2003 EL61 and use these observations to set tight constraints on its shape, density ($\rho$), size, and visual albedo ($p_v$). Since the body is very large, the shape it takes is largely determined by self-gravity and centripetal forces. Owing to the high gravitational pressure, the body will deform like a viscous fluid and relax over time to the equilibrium figure of a rotating, homogenous body. For example, icy bodies with radii ~1000 km have central pressures ~$10^8$ Pa (Lewis 1997). With differential stresses of this magnitude and at temperatures ~140 deg K, water ice deforms like a fluid at strain rate ~$10^{-7}$ s$^{-1}$ (Durham, Kirby, & Stern 1997). Rocky materials behave in a similar way at these high pressures (Kohlstedt, Evans, & Mackwell 1995). A 1000-km scale body will therefore relax to an equilibrium shape on a time scale ~(strain rate)$^{-1}$ or ~100 days. Following Farinella. Paolicchi, & Zappala (1981) and Sheppard & Jewitt (2002) who analyze the shapes of asteroids and smaller KBOs assuming they are strengthless rubble piles, we determine the shape and density of 2003 EL61 from the rotation period and light curve amplitude. In our case, however, we assume a body of nominal strength, yet viscous on long time scales owing to self-gravity. Since we know the mass from the satellite orbit, we are also able to determine the size and albedo.

2. OBSERVATIONS

Most of the observations we report here were acquired 2005 Jan 25 to 2005 Jul 26 by an on-site operator at Cerro Tololo using the SMARTS 1.3m telescope and the optical channel of ANDICAM camera which has dual IR/optical CCDs (Bloom et al. 2004). The optical channel is a Fairchild 2Kx2K CCD which we used in 2x2-binned mode to obtain 0.34" per pixel and a 5.8' x 5.8' field of view. This telescope is queue-scheduled for shared use by all members of the SMARTS consortium. The CCD camera is permanently mounted, and users share sky flats, bias frames, and observations of Landolt standards taken in four Johnson filters (B, V, R, and I) at varying airmass on photometric nights. All exposures are auto-guided and the typical seeing is 1". Our nightly observations of 2003 EL61 were four consecutive exposures in B, V, and twice in I or else two sequences in B, V, R, and I. All exposures were 300 s for B and 120 s for V, R, and I.

To obtain a measure of 2003 EL61's rotational light curve unbiased by the 24-hour sampling interval of our SMARTS observations, we made additional observations spanning the nights of 2005 Apr 10, 11, and 12 with the 200" Hale telescope at Palomar Observatory and the night of 2005 May 4 with the 32" Telescope at Tenagra Observatory. At Palomar, we used the Large Format Camera, a CCD mosaic with a field size is of 24.6 arcminutes made from six 2Kx4K CCDs. All the observations are 2 minute exposures through a Gunn r' filter. Seeing was 1.25 to 1.8 arcsec and variable. At Tenagra, we used the facility 1Kx1K CCD camera and Johnson R filter.

To reduce the SMARTS data, we first used nightly bias frames, darks, and twilight flats to flatten the images. We then used observations of 2003 EL61 and of Landolt stars taken on the same photometric nights to determine the apparent magnitudes in the Johnson system of hand-selected field stars surrounding 2003 EL61. With these secondary standards calibrated, we were then able to calibrate the magnitudes we measured for 2003 EL61 on both photometric and non-photometric nights. We applied standard aperture-correction techniques to minimize the size of the measurement aperture, relying upon the field stars in each image to determine the aperture correction. For all the SMARTS observations we report here, we used a small aperture with a diameter of 6 pixels (2.0") for 2003 EL61, and a large aperture with a diameter of 40 pixels (13.6") for Landolt standards. We also determined color corrections from our observations of Landolt stars on photometric nights. These corrections were insignificant for 2003 EL61, except for the B-band measurements where an 0.013 magnitude correction was required.

We judged some of the SMARTS observations to be unusable owing to very large errors. Most of these were either taken during full moon conditions or very poor conditions. A few observations were adversely affected by cosmic ray hits within the flux aperture. We do not include these observations in our results. We also rejected observations in each filter with measurement errors exceeding 0.3 mags, or with magnitudes differing by more than three times the root-mean-square variation of the total sample. Additional details of our measurement procedure appear in previous publications (Schaefer & Rabinowitz 2002, Schaefer & Tourtellotte 2001).

To reduce the Palomar and Tenagra observations, we first dark-subtracted and flat-fielded each image. Then, choosing a fixed aperture, we measured the sky-subtracted flux of 2003 EL61 and of a fixed set of field stars selected from the USNO catalog. There was one fixed set for Palomar and another for Tenagra. The nights were not photometric and we made no attempt to calibrate with respect to standards. Instead, we normalized each brightness measurement of 2003 EL61 with respect to the mean brightness of the field stars in the same exposure. The magnitudes we report here for 2003 EL61 are scaled so that the mean Palomar and Tenagra magnitudes match the mean V-band measurement from the SMARTS observations.

## 3. RESULTS AND ANALYSIS

### 3.1 Light curve data

Table 1 lists the Julian date, apparent and reduced magnitudes, magnitude errors, solar phase angle ($\alpha$), heliocentric distance (r), geocentric distance (d), filter, and the telescope we used for each observation of 2003 EL61. The reduce magnitude is the apparent magnitude minus 5log(r d) where r and d are in AU. Extrapolated to $\alpha$=0 deg, the reduced magnitude yields the absolute magnitude of 2003 EL61 in each pass band. The magnitude error is the statistical error we determine from the read noise and from counting the photoelectrons from the sky and from 2003 EL61 within our measurement aperture. We also include a random error of 0.015 mags. This accounts for night-to-night variations in the magnitudes we measure for field stars which is not otherwise accounted for by our statistical estimate. The observations we have excluded as outliers (see discussion above) are marked "rejected". The Palomar and Tenagra observations are marked "relative" because they are not calibrated to a standard magnitude system owing to cirrus clouds during the time of the observations.

We have also adjusted the Julian dates listed in Table I by subtracting the relative light-travel time of the observations, $(d-d_0)/c$, where d is the geocentric distance of 2003 EL61 at the time of the observation, $d_0$ is the distance at the time of our first SMARTS observation in Table I, and c is the speed of light. For observations that span only a few nights, this correction would be insignificant. Given the long time span of our observations, however, this correction turns out to be marginally important (at the few percent level for rotational phase).

3.2 Rotational Light curve

Figure 1 shows the measurements from Table I plotted versus Julian date for each filter. There is clearly a scatter in the observations that is much larger than their errors. At this point we can already calculate average magnitudes for each data set, and linear fits to magnitude versus solar phase angle. Because we have many observations spanning a long time interval the scatter largely cancels. However, these derivations are preliminary and we do not report them here. We make these derivations only for the purpose of combining all the data plotted in Figure 1 into one set for which there is no dependence on solar phase angle or on wavelength. From this set we can best determine the rotational light curve. We then subtract the rotation curve from the original data plotted in Figure 1 and re-derive the average magnitudes, colors, and fits to the solar phase curves. The method is iterative, but leads to a unique solution after only one iteration. Note that we only fit the solar phase curves for the SMARTS, B, V, and I data. The solar phase angle does not vary significantly for the dates spanned by the Palomar and Tenagra data sets. We also omit the R-band band data because they are few in number and span only a few nights.

In Figure 2, we show a periodogram for the combined data, computed using the phase-dispersion minimization (pdm) program distributed with the IRAF image-processing software (Stellingwerf 1978). Here the vertical axis, $\theta$, represents the dispersion with respect to the mean rotational light curve computed for period p. More precisely, if the data consists of magnitudes $M_i$ at times $t_i$, with i = 1 to N, then each observation has rotational phase, $\phi_i$, = $[t_i$ modulo p]/p. Binning the observations by phase and averaging the values of $M_i$ within each bin yields the mean rotational light curve, $F(\phi)$. The dispersion with respect to $F(\phi)$ is the sum $\Sigma_{i=1,N} (M_i - F(\phi_i))^2 / (N-1)$. Dividing this sum by the dispersion with respect to the overall mean, $\Sigma_{i=1,N} (M_i - <M>)^2 / (N-1)$, yields $\theta$. The best fit periods are those at the smallest values of $\theta$.

Two clear minima for $\theta$ appear near periods of 0.08 and 0.16 days in Figure 2. We also looked for minima at periods from 0.2 to 100 days and did not find any of any significance other than integer multiples of these fundamental periods. From inspection of the dispersion plot at higher resolution, and from inspection of the rotational light curves associated with each minimum, we find the best-fit periods of 0.08157 ± 0.00001 days (1.9577 ± 0.002 h) for a single-peaked light curve and 0.163135 ± 0.00001 days (3.9154 ±0.0002 h) for a double-peaked light curve.

Figure 3 shows the composite data plotted versus rotational phase for the 3.9154-h period. A different symbol represents each of the data sets going into the composite. Also plotted is a fit made by binning the data by rotational phase, with bin widths equal to 1/40th of the period, and computing the median average of the composite data within each bin. The curve was then smoothed by replacing each point along the curve with the average of itself and its two closest neighbors. We derive an overall 2% error to the fit, which is the rms scatter of the observations within each bin divide by the square root of the number of observations per bin.

Note that the fit in Figure 3 has two peaks with respective amplitudes of 0.08 and 0.15 mags and two valleys with depths of 0.08 and 0.13 mags. The chi-square of the composite data with respect to this fit is 583 for 306 degrees of freedom (after rejecting 3-sigma outliers), which indicates that there is still a lot of scatter after removal of the rotation. Repeating the fitting procedure for the 1.9577-h period, we find a larger chi-square of 640 with 297 degrees of freedom. Because the longer period yields a better chi-square fit, and because the longer-period light curve consists of two half-cycles with different amplitudes, we conclude that the longer period is the true period. In any case, the rotation period for the single-peaked curve is so short that a strengthless body would require $\rho > 10{,}240$ kg m$^{-3}$ to stay together (Chandrasekhar 1969). Even a body with the strength typical of terrestrial soils would require $\rho > 4100$ kg m$^{-3}$ (Holsapple 2001).

To look for any dependence on wavelength, we plot the SMARTS B, V, and I observations versus rotational phase (Fig. 4) after binning by rotational phase (bin size equal to the period/20). Here each point shows the median average of the magnitude within each bin. Where the number of observations (n) per bin is two or more, the error bar show the root-mean-square deviation from the median divided by the square root of n. Where n=1, the error bar is the estimated error of the observation. Figure 4 also shows the fit to the composite phase curve from Figure 3.

Examination of Figure 4 shows there are only a few data points that differ from the composite fit by more than their errors. But these deviations are not consistent near the peaks or the valleys in the light curve. The most significant deviation is a single point in the V-band, falling ~0.1 mag below the crest of the larger of the two peaks (rotational phase 0.62). But this point corresponds to a bin for which there was only one observation (see Figure 3). Also, there is no corresponding deviation in the I-band at the same rotational phase, which we would expect if the reflectance spectrum where changing significantly there. For these reasons, we conclude that there is no significant wavelength dependence to the rotation curve at the few percent level or larger.

3.3 Solar Phase Curves

To determine the solar phase curves, we first subtracted our fit to the composite rotational light curve (see Figure 3) from the original, uncombined data (see Figure 1). We did this by calculating the rotational phase at the light-travel-corrected time of each observation. We then evaluated the amplitude of the rotation curve corresponding to each observation. Since we computed the rotation curve for 40 bins in rotational phase, we determined the amplitude for each observation by linearly extrapolating the curve from the two bins closest to the rotational phase of the observation. With the rotation curve subtracted, we then separately sorted the SMARTS B, V, and I observations by solar phase angle, and computed the weighted averaged of the reduced magnitude every 10 consecutive data points.

Figure 5 shows the average reduced magnitude versus solar phase angle for these rotation-corrected observations. Also shown are linear fits for each filter. It is clear that the phase curves are linear over the full range in solar phase of the observations (0.5 to 1.1 deg). To fit the observations, we gave each point in Figure 5 equal weight. The resulting slope and intercept (parameters a and b, respectively) are listed in Table II for each filter. In each case we calculated the errors for a and b for each assuming the fitted points had the same magnitude error, equal to the standard deviation of the residual magnitudes after subtraction of the fit. Note that parameter b is the absolute magnitude for 2003 EL61 in each filter, which is the reduced magnitude extrapolated to zero phase angle.

3.4 Average Magnitudes and Colors

Table 3 lists the rotation-corrected, reduced B, V, R, I magnitudes and their differences that we measured with the SMARTS telescope the nights of 2005 Jan 25, 26, and 27. On each of these nights we recorded two sequences in B, V, R, and I, except for the last night where we had to reject our second B-band measurement. Here we have subtracted our fit to the rotational light curve from each observation using the method discussed above. This allows us to determine the B-V, V-R, R-I and V-I colors without the influence of rotational modulation. Table 3 also lists the weighted average values of the magnitudes and colors from all three nights. Subtracting solar colors (B-V = 0.67, V-R = 0.36, V-I = 0.69 from Delsanti et al. 2001) from the average magnitude differences for 2003 EL61, we obtain reflectance colors B-V = -0.044 ± 0.025, V-R = -0.017 ± 0.020 and V-I = -0.007 ± -0.020. These values being close to or consistent with zero, the reflectance for 2003 EL61 is neutral to within a few percent. The slightly negative value for the B-V color, while marginally significant, suggests a lightly enhanced reflectance at blue wavelengths.

4. DISCUSSION

4.1 color and intrinsic brightness

The neutral color we observe for 2003 EL61 is typical of many smaller KBOs, and in this respect unremarkable. As with the smaller bodies, the lack of color is consistent with a surface consisting of dark carbonaceous material mixed with ices. Depending on the amount of ice and how it is mixed in with darker material, the albedo could be very low (<0.05) or very high (>0.5). Since the near infrared spectrum for 2003 EL61 shows the clear presence of water ice, the albedo could be very high if the darker material is hidden beneath the ice.

 In Figure 6, we compare the sun-subtracted value of V-I we observe for 2003 EL61 with the values known for all the intrinsically bright TNOs (absolute magnitude < 4). Here we see that a relatively neutral color is a common feature of the intrinsically brightest TNOs. With the exception of Sedna, all of the TNOs with absolute magnitudes less than 2.5 (2003 UB13, Pluto, 2005 FY9, 2003 EL61, Charon, and Orcus) have V-I < 0.2, whereas the relatively fainter bodies (Quaoar, Ixion, 2002 AW197, and Varuna) have V-I = 0.3 to 0.6. This higher range is also representative of the whole TNO population, for which the average solar-corrected values for V-I range from 0.36 to 0.50 depending on the subgroup (Hainaut and Delsanti 2002).
We speculate that this correlation, if real, is evidence that the ice-covered surfaces on these largest bodies are covering up the redder, darker areas underneath, as we suggest above for the case of 2003 UB313.

4.2 solar phase curve

The linear solar phase curve we observe for 2003 EL61 is also typical of many other TNOs (Schaefer, Rabinowitz, & Tourtellotte 2005), although the slope we measure for the B and V band (~0.09 mag deg$^{-1}$) is shallower than average (~0.15 mag deg$^{-1}$, Sheppard and Jewitt 2002). Pluto also has a very shallow phase curve (Buratti et al 2003), although the reason for this is not well understood. Opposition surges depend in a complex way on the optical properties of surface grains, their size distribution, and their spatial distribution. The amplitude and width of the surge depend on shadow hiding and on multiple scattering by surface grains. In general, the slopes of phase curve at low phase angles are not directly correlated with albedo (Schaefer & Rabinowitz 2002). That we see shallow phase curves for two of the largest known TNOs could be an indication of similar optical properties or surface structure.

For 2003 EL61, we also see a wavelength dependence to the slope of the phase curve, with the longer wavelengths having a steeper slope. Recently, we have observed this same dependence for smaller TNOs (Schaefer, Rabinowitz, & Tourtellotte 2005). It might be that the red albedo for these objects is high enough to allow light to coherently back scatter, whereas at blue wavelengths the albedo is too low for multiple scattering to dominate shadow hiding (Rabinowitz, Schaefer, & Tourtellotte 2004). However, we show below that 2003 EL61 must have a very high albedo in the blue as well as the red wavelengths. So there may be more subtle effects occurring, such as a grain structure that enhances the scattering of red light relative to blue.

4.3 Size, shape, density, and albedo

The rotation period we measure for 2003 EL61 is extraordinary. It is the shortest rotation reliably measured for any TNO (Sheppard & Jewitt 2002, 2004), and for any solar system body larger than 100 km (Pravec, Harris, & Michalowski 2002). Even a solid body of moderate strength would be significantly distorted at this high rotation rate (Holsapple 2001). Assuming 2003 EL61 responds to stresses like a fluid, with the finite viscosity controlling the rate of deformation, the high rotation speed requires that 2003 EL61 eventually take the shape of either an oblate (Maclaurin) spheroid or else that of a triaxial (Jacobi) ellipsoid (Chandrasekhar 1969). Either shape is possible depending on the density and rotation period. For the case of the Jacobi ellipsoid, the variation in the projected surface area produces the double-peaked light curve. The shape and density are determined by the rotation period and light curve amplitude using the tables provided by Chandrasekhar (Farinella et al. 1981, Jewitt and Sheppard 2002).

The Maclaurin spheroid, on the other hand, does not produce a light curve because of its rotational symmetry. To produce the double-peaked light curve, there has to be a peculiar albedo pattern across the surface, consisting of two albedo spots on opposite sides of the equator. We discuss this further below. Given the rotation period of 3.9154 h, Chandrasekhar shows that dynamical stability for a Maclaurin spheroid requires $\rho > 2530$ kg m$^{-3}$. At lower densities, only the ellipsoidal shape is stable. Since we expect bodies in the Kuiper Belt to be composed of mixtures of ice and rock with $\rho < 3300$ kg m$^{-3}$ (the mean density of Earth's moon, which is all rock and no ice), the range of possible densities is very narrow. Knowing the rotation period, we again use Chandrasekhar's tables to constrain the shape.

First let us determine the shape constraints for the Jacobi ellipsoid. Let the ellipsoid's three axes have lengths $a_1 > a_2 > a_3$, where axis three is the rotation axis. Let $\phi$ be the angle between the rotation axis and our line of site. The peak-to-peak amplitude we measure for 2003 EL61's light curve ($\Delta m = 0.28 \pm 0.04$ mags) gives a lower bound $a_1/a_2 > 10^{0.4*\Delta m} = 1.29 \pm 0.04$, with the equality realized when $\phi = 90°$. Given our measured period of 3.9154 h, we can then use Chandrasekhar's table relating $a_1/a_2$, $a_1/a_3$, $\rho$, and angular velocity for Jacobi ellipsoids to obtain $a_1/a_3$ and $\rho$. Using $m = 4\pi a_1 a_2 a_3 \rho / 3 = 4.21 \pm 0.1 \times 10^{21}$ kg, we then determine $a_1$, $a_2$, and $a_3$. Finally from our measurement of 2003 EL61's absolute visual magnitude ($m_v = 0.44$, see table 2), we determine the visual albedo, $p_v$ using

$$p_v = 2.25 \times 10^{16}/(a_3[a_1 a_2]^{1/2}) \times 10^{(m_s - m_v)/2.5}$$

where $m_s$ is the apparent visual magnitude of the sun (-26.7) and $a_3[a_1 a_2]^{1/2}$ is the square of the mean effective radius (see Sheppard and Jewitt 2004).

Table 4 shows the resulting values for $\rho$, $a_1$, $a_2$, and $p_v$ for $a_1/a_2$ ranging from 1.316 up to 2.314. Note that these are all the values for $a_1/a_2$ tabulated by Chandrasekhar exceeding our lower bound, yet lower than the critical value, $a_1/a_2 = 2.314$, at which the homogenous ellipsoid becomes unstable. For each value of $a_1/a_2$, the table also lists the associated value for $\phi$ determined from the dependence of $\Delta m$ on $\phi$, $a_1$, $a_2$, and $a_3$ (Binzel et al. 1989):

$$\Delta m = 2.5[\log(a_1/a_2) - \log(r_1/r_2)],$$

where
$$r_1 = [a_1^2 \cos^2(\phi) + a_3^2 \sin^2(\phi)]^{1/2}$$
and
$$r_2 = [a_2^2 \cos^2(\phi) + a_3^2 \sin^2(\phi)]^{1/2}$$

Note that $[r_1 r_2]^{1/2}$ replaces $a_3$ in the above equation for $p_v$ when $\phi < 90°$. Extrapolating the values listed in table 4 to our lower bound $a_1/a_2 = 1.29 \pm 0.02$, we find minimum values $\rho > 2600 \pm 37$ kg m$^{-3}$, $a_1 > 980 \pm 37$ km and maximum values $p_v < 0.73 \pm 0.01$, $a_2 < 759 \pm 24$ km, and $a_3 < 498 \pm 4$ km. Since it is likely that 2003 EL61's rotation axis is coincident with the orbital plane of its satellite, which we know to be inclined by only 4° from our line of site (Brown et al 2005), we can assume that 2003 EL61's rotation axis makes an angle $\phi = 86°$ with our line of site. Correcting for the projection increases $a_1/a_2$ by only 0.2%, so we can take the above limiting values as actual values.

In the unlikely case that we are viewing the rotation axis closer to pole, then we can assume the upper stability limit $a_1/a_2 = 2.314$ which occurs at $\phi = 47°$. In that case we obtain upper limits $\rho < 3340$ kg m$^{-3}$, $a_1 < 1250$ km, and lower limits $p_v > 0.60$, $a_2 > 540$ km, and $a_3 > 430$ km. Remarkably, the resulting solutions impose a very limited range of variation for $a_3$ (7%), with somewhat large ranges for $p_v$ (9%), $\rho$ (12 %), $a_1$ (12 %), and $a_2$ (17 %). These values are summarized in Table 5.

For the case of the Maclaurin spheroid spinning with period 3.9154 h, let $a_1$ and $a_3$ be the radii to the equator and to the pole, respectively. Chandrasekhar's tables give ratio $a_1/a_3$ = 1.72 at the limit of stability where $\rho$ = 2530 kg m$^{-3}$ and give $a_1/a_3$=1.42 if the density is as high as for Earth's moon, $\rho$= 3300 kg m$^{-3}$. Given the known mass and the measured brightness, we thus calculate the ranges for $a_1$, $a_3$, and $p_v$ listed in Table 5. They are comparable to the ranges we obtained for the Jacobi ellipsoid. In either case, we end up with a squashed figure, $\rho \sim$ 2500 kg m$^{-3}$ and a short axis ~500 km wide, although the spheroid requires somewhat higher albedo in the range 0.7 to 0.8 and an overall length of ~1600 km, shorter than the longest dimension of the ellipsoid by 15%. On this basis alone, there is no reason to choose one model over the other.

The only reason to exclude the Maclaurin spheroid is because of the special albedo pattern that would be required to mimic the double-peaked light curve. Pluto, which rotates too slowly to deform significantly, has albedo patterns which account for the light curve shape (Stern, Buie, & Trafton 1997). The light curve variations are of comparable amplitude to what we see for 2003 EL61, but it is a single-peaked light curve. We do not see significant color variation above a few percent level for 2003 EL61, which we might expect from the difference in composition between the bright and dark areas. But since Pluto's disk-averaged color only varies by ~1% (Buratti et al. 2003) color variability is not necessarily diagnostic. Hence we have no observational basis to exclude the possibility of an oblate spheroid with bright spots on opposite hemispheres.

Another possibility is that 2003 EL61 is a binary (making 2003 EL61 a tertiary system when we include the co-orbiting satellite). In this case the mutual eclipses of the close, co-orbiting pair cause the light curve variations. But Leone et al (1984) show that such a binary configuration is unlikely if the light curve amplitude is small and the rotational velocity is high, as is the case for 2003 EL61. They tabulate approximate equilibrium solutions assuming the co-orbiting bodies are homogenous and strengthless, but of unequal mass. In this case each body takes the shape of a triaxial ellipsoid distorted by its own rotation and by the gravity of the other body. With these assumptions, and given the short rotation period we observe, there is no stable solution for $\rho$ < 5000 kg m$^{-3}$. This clearly rules out a contact binary.

5. CONCLUSIONS

We have shown that 2003 EL61 is a neutral colored body with linear solar phase curve, the slope of which is wavelength dependent. Comparing 2003 EL61 to other TNOs of comparable intrinsic brightness, we see that these largest bodies are all relatively neutral in color compared to their smaller cousins. As for the case of 2003 UB313, we suggest that 2003 EL61 and these largest Kuiper Belt objects have relatively neutral colors because their surfaces are covered with frozen methane or water, hiding the presence of darker, redder material underneath.

The very short rotation period for 2003 EL61 is extraordinary, and provides us the first opportunity to study the extreme rotational deformation of large body for which gravitational pressure exceeds strength. If the body has relaxed to the figure expected for a fluid body, then from the amplitude of the rotational light curve and the rotation period we accurately determine the shape and density. Given the mass determined for the satellite orbit, we also tightly constrain the size and albedo. We present two cases, one assuming 2003 EL61 is an elongated ellipsoid whose shape determines the light curve variation, and the other a flattened spheroid for which albedo patterns cause the variation. Both cases yield similar results for the shape, size, density, and albedo, but with the ellipsoid being slightly more elongated (at least 1960 km long) and slightly darker ($p_v$ = 0.6 to 0.7) compared to the spheroid (1600 km maximum width, $p_v$ = 0.7 to 0.8). We believe the spheroidal model to be much less likely, however, given the peculiar pattern of albedo spots that would be required to produce a double-peaked light curve.

Forthcoming observations in the thermal infrared with the Spitzer space telescope will independently determine the albedo. Also, direct imaging at visible wavelengths with the Hubble Space Telescope may constrain the size and shape. Confirmation of our size, shape, and albedo estimates would then be strong evidence in support of the deformation model. A more detailed ground based study of the rotational light curve may further constrain the shape and distinguish between the two solutions.

ACKNOWLEDGEMENTS

This work was supported by grants from NASA Planetary Astronomy.## 6. REFERENCE

Binzel, R. P., Farinella, P., Zappala, V., & Cellino, A. 1989, In Asteroids II, eds. R. P. Binzel, T. Gehrels, & M. S. Matthews (Tucson: University of Arizona Press), 416

Bloom, J. S., van Dokkum, P. G., Bailyn, C. D., Buxton, M. M., Kulkarni, S. R.;,Schmidt, B. P. 2004, AJ 127, 252

Boehnhardt, H., Bagnulo, S., Muinonen, K., Barucci, M. A., Kolokolova, L., Dotto, E., & Tozzi, G. P. 2004, A&A 415, L21.

Brown, M.E., Trujillo, C. A., & Rabinowitz, D. L. 2004, ApJ 617, 645

Brown, M.E., Trujillo, C. A., & Rabinowitz, D. L. 2005, ApJL, in press

Brown, M. E., Bouchez, A. H., Rabinowitz, D., Sari, R., Trujillo, C. A, van Dam, M., Campbell, R., Chin, J., Hartman, S., Johansson, E., Lafon, R., LeMignant, D., Stomski, P., Summers, D., & Wizinowich, P. 2005, ApJ 632, L45

Buratti, B. J., Hillier, J. K., Heinze, A., Hicks, M. D., Tryka, K. A., Mosher, J. A., Ward, J., Garske, M., Young, J., & Atienza-Rosel, J. 2003, Icarus 162, 17111

Table 1. Measured reduced magnitudes for 2003 EL61

| JD-2450000 (corrected) | Apparent Mag | Reduced Mag | Mag Err | α (Deg) | r (AU) | d (AU) | Filter | Telescope | Notes |
|---|---|---|---|---|---|---|---|---|---|
| 3396.85297 | 18.133 | 1.051 | 0.036 | 1.024 | 51.255 | 50.892 | B | S-1.3m | |
| 3396.85609 | 17.466 | 0.384 | 0.042 | 1.024 | 51.255 | 50.892 | V | S-1.3m | |
| 3396.85816 | 17.183 | 0.101 | 0.028 | 1.024 | 51.255 | 50.892 | R | S-1.3m | |
| 3396.86028 | 16.876 | -0.206 | 0.023 | 1.024 | 51.255 | 50.892 | I | S-1.3m | |
| 3396.86344 | 18.247 | 1.165 | 0.028 | 1.024 | 51.255 | 50.892 | B | S-1.3m | |
| 3396.86656 | 17.668 | 0.586 | 0.046 | 1.024 | 51.255 | 50.892 | V | S-1.3m | |
| 3396.86862 | 17.268 | 0.186 | 0.039 | 1.024 | 51.255 | 50.892 | R | S-1.3m | |
| 3396.87073 | 16.968 | -0.114 | 0.023 | 1.024 | 51.255 | 50.892 | I | S-1.3m | |
| 3397.83262 | 18.089 | 1.008 | 0.027 | 1.018 | 51.255 | 50.878 | B | S-1.3m | |
| 3397.83573 | 17.532 | 0.451 | 0.047 | 1.018 | 51.255 | 50.878 | V | S-1.3m | |
| 3397.83780 | 17.157 | 0.076 | 0.037 | 1.018 | 51.255 | 50.878 | R | S-1.3m | |
| 3397.83993 | 16.896 | -0.185 | 0.024 | 1.018 | 51.255 | 50.878 | I | S-1.3m | |
| 3397.84309 | 18.293 | 1.212 | 0.096 | 1.018 | 51.255 | 50.878 | B | S-1.3m | |
| 3397.84620 | 17.534 | 0.453 | 0.037 | 1.018 | 51.255 | 50.878 | V | S-1.3m | |
| 3397.84827 | 17.213 | 0.132 | 0.023 | 1.018 | 51.255 | 50.878 | R | S-1.3m | |
| 3397.85040 | 16.984 | -0.097 | 0.026 | 1.018 | 51.255 | 50.878 | I | S-1.3m | |
| 3398.82554 | 18.258 | 1.177 | 0.027 | 1.011 | 51.254 | 50.864 | B | S-1.3m | |
| 3398.82865 | 17.669 | 0.588 | 0.032 | 1.011 | 51.254 | 50.864 | V | S-1.3m | |
| 3398.83072 | 17.272 | 0.191 | 0.029 | 1.011 | 51.254 | 50.864 | R | S-1.3m | |
| 3398.83284 | 16.896 | -0.185 | 0.031 | 1.011 | 51.254 | 50.864 | I | S-1.3m | |
| 3398.83911 | 17.605 | 0.524 | 0.037 | 1.011 | 51.254 | 50.864 | V | S-1.3m | |
| 3398.84118 | 17.349 | 0.268 | 0.023 | 1.011 | 51.254 | 50.864 | R | S-1.3m | |
| 3398.84330 | 16.890 | -0.191 | 0.028 | 1.011 | 51.254 | 50.864 | I | S-1.3m | |
| 3401.82456 | 18.080 | 1.001 | 0.173 | 0.991 | 51.254 | 50.822 | B | S-1.3m | |
| 3401.82767 | 17.349 | 0.270 | 0.132 | 0.991 | 51.254 | 50.822 | V | S-1.3m | |
| 3401.82977 | 16.870 | -0.209 | 0.133 | 0.991 | 51.254 | 50.822 | I | S-1.3m | |
| 3401.83181 | 16.811 | -0.268 | 0.105 | 0.991 | 51.254 | 50.822 | I | S-1.3m | |
| 3403.83751 | 18.149 | 1.071 | 0.026 | 0.976 | 51.254 | 50.795 | B | S-1.3m | |
| 3403.84062 | 17.527 | 0.449 | 0.020 | 0.976 | 51.254 | 50.795 | V | S-1.3m | |
| 3403.84273 | 16.842 | -0.236 | 0.022 | 0.976 | 51.254 | 50.795 | I | S-1.3m | |
| 3403.84476 | 16.815 | -0.263 | 0.023 | 0.976 | 51.254 | 50.795 | I | S-1.3m | |
| 3404.85250 | 18.197 | 1.120 | 0.029 | 0.968 | 51.254 | 50.781 | B | S-1.3m | |
| 3404.85562 | 17.637 | 0.560 | 0.027 | 0.968 | 51.254 | 50.781 | V | S-1.3m | |
| 3404.85773 | 16.944 | -0.133 | 0.024 | 0.968 | 51.254 | 50.781 | I | S-1.3m | |
| 3404.85976 | 16.960 | -0.117 | 0.024 | 0.968 | 51.254 | 50.781 | I | S-1.3m | |
| 3405.82495 | 18.143 | 1.066 | 0.018 | 0.960 | 51.254 | 50.769 | B | S-1.3m | |
| 3405.82806 | 17.551 | 0.474 | 0.019 | 0.960 | 51.254 | 50.769 | V | S-1.3m | |
| 3405.83016 | 16.909 | -0.168 | 0.020 | 0.960 | 51.254 | 50.769 | I | S-1.3m | |
| 3405.83219 | 16.940 | -0.137 | 0.021 | 0.960 | 51.254 | 50.769 | I | S-1.3m | |
| 3407.83966 | 18.218 | 1.143 | 0.023 | 0.943 | 51.253 | 50.743 | B | S-1.3m | |
| 3407.84277 | 17.549 | 0.474 | 0.021 | 0.943 | 51.253 | 50.743 | V | S-1.3m | |
| 3407.84487 | 16.858 | -0.217 | 0.041 | 0.943 | 51.253 | 50.743 | I | S-1.3m | |
| 3407.84691 | 16.868 | -0.207 | 0.055 | 0.943 | 51.253 | 50.743 | I | S-1.3m | |
| 3408.82750 | 18.159 | 1.084 | 0.018 | 0.935 | 51.253 | 50.730 | B | S-1.3m | |

| | | | | | | | | |
|---|---|---|---|---|---|---|---|---|
| 3408.83061 | 17.535 | 0.460 | 0.019 | 0.935 | 51.253 | 50.730 | V | S-1.3m |
| 3408.83271 | 16.881 | -0.194 | 0.020 | 0.935 | 51.253 | 50.730 | I | S-1.3m |
| 3408.83475 | 16.875 | -0.200 | 0.020 | 0.935 | 51.253 | 50.730 | I | S-1.3m |
| 3409.81989 | 18.181 | 1.107 | 0.018 | 0.926 | 51.253 | 50.717 | B | S-1.3m |
| 3409.82300 | 17.583 | 0.509 | 0.020 | 0.926 | 51.253 | 50.717 | V | S-1.3m |
| 3409.82510 | 16.936 | -0.138 | 0.021 | 0.926 | 51.253 | 50.717 | I | S-1.3m |
| 3409.82714 | 16.969 | -0.105 | 0.021 | 0.926 | 51.253 | 50.717 | I | S-1.3m |
| 3410.84631 | 18.266 | 1.192 | 0.019 | 0.917 | 51.253 | 50.705 | B | S-1.3m |
| 3410.84942 | 17.618 | 0.544 | 0.019 | 0.917 | 51.253 | 50.704 | V | S-1.3m |
| 3410.85152 | 16.931 | -0.143 | 0.020 | 0.917 | 51.253 | 50.704 | I | S-1.3m |
| 3410.85354 | 16.884 | -0.190 | 0.020 | 0.917 | 51.253 | 50.704 | I | S-1.3m |
| 3411.85775 | 18.133 | 1.060 | 0.017 | 0.908 | 51.253 | 50.692 | B | S-1.3m |
| 3411.86086 | 17.533 | 0.460 | 0.019 | 0.908 | 51.253 | 50.692 | V | S-1.3m |
| 3411.86499 | 16.912 | -0.161 | 0.023 | 0.908 | 51.253 | 50.692 | I | S-1.3m |
| 3412.81093 | 17.584 | 0.511 | 0.020 | 0.899 | 51.253 | 50.680 | V | S-1.3m |
| 3412.81303 | 16.908 | -0.165 | 0.021 | 0.899 | 51.253 | 50.680 | I | S-1.3m |
| 3412.81506 | 16.891 | -0.182 | 0.022 | 0.899 | 51.253 | 50.680 | I | S-1.3m |
| 3413.84029 | 18.378 | 1.306 | 0.022 | 0.890 | 51.253 | 50.668 | B | S-1.3m |
| 3413.84340 | 17.782 | 0.710 | 0.044 | 0.890 | 51.253 | 50.668 | V | S-1.3m |
| 3413.84551 | 17.101 | 0.029 | 0.023 | 0.890 | 51.253 | 50.668 | I | S-1.3m |
| 3413.84755 | 17.106 | 0.034 | 0.030 | 0.890 | 51.253 | 50.668 | I | S-1.3m |
| 3414.85838 | 18.238 | 1.166 | 0.021 | 0.880 | 51.253 | 50.656 | B | S-1.3m |
| 3414.86149 | 17.545 | 0.473 | 0.023 | 0.880 | 51.253 | 50.656 | V | S-1.3m |
| 3414.86359 | 16.889 | -0.183 | 0.026 | 0.880 | 51.253 | 50.656 | I | S-1.3m |
| 3420.87172 | 18.281 | 1.212 | 0.069 | 0.820 | 51.252 | 50.590 | B | S-1.3m |
| 3420.87483 | 17.613 | 0.544 | 0.041 | 0.820 | 51.252 | 50.590 | V | S-1.3m |
| 3420.87693 | 16.948 | -0.121 | 0.053 | 0.820 | 51.252 | 50.590 | I | S-1.3m |
| 3421.85236 | 18.272 | 1.204 | 0.023 | 0.810 | 51.252 | 50.579 | B | S-1.3m |
| 3421.85546 | 17.626 | 0.558 | 0.027 | 0.810 | 51.252 | 50.579 | V | S-1.3m |
| 3421.85756 | 16.939 | -0.129 | 0.022 | 0.810 | 51.252 | 50.579 | I | S-1.3m |
| 3421.85958 | 16.863 | -0.205 | 0.025 | 0.810 | 51.252 | 50.579 | I | S-1.3m |
| 3422.79518 | 18.183 | 1.115 | 0.019 | 0.801 | 51.252 | 50.570 | B | S-1.3m |
| 3422.79829 | 17.612 | 0.544 | 0.022 | 0.801 | 51.252 | 50.570 | V | S-1.3m |
| 3422.80040 | 16.924 | -0.144 | 0.027 | 0.801 | 51.252 | 50.570 | I | S-1.3m |
| 3422.80242 | 16.929 | -0.139 | 0.023 | 0.801 | 51.252 | 50.570 | I | S-1.3m |
| 3423.78341 | 18.336 | 1.268 | 0.037 | 0.790 | 51.252 | 50.560 | B | S-1.3m |
| 3423.78651 | 17.639 | 0.571 | 0.041 | 0.790 | 51.252 | 50.560 | V | S-1.3m |
| 3423.78862 | 16.992 | -0.076 | 0.047 | 0.790 | 51.252 | 50.560 | I | S-1.3m |
| 3423.79065 | 16.869 | -0.199 | 0.056 | 0.790 | 51.252 | 50.560 | I | S-1.3m |
| 3424.78988 | 18.253 | 1.186 | 0.029 | 0.780 | 51.252 | 50.550 | B | S-1.3m |
| 3424.79299 | 17.632 | 0.565 | 0.029 | 0.780 | 51.252 | 50.550 | V | S-1.3m |
| 3424.79510 | 16.902 | -0.165 | 0.044 | 0.780 | 51.252 | 50.550 | I | S-1.3m |
| 3424.79713 | 16.912 | -0.155 | 0.026 | 0.780 | 51.252 | 50.550 | I | S-1.3m |
| 3425.81299 | 18.174 | 1.107 | 0.037 | 0.769 | 51.252 | 50.541 | B | S-1.3m |
| 3425.81610 | 17.609 | 0.542 | 0.038 | 0.769 | 51.252 | 50.541 | V | S-1.3m |
| 3425.81821 | 16.935 | -0.132 | 0.027 | 0.769 | 51.252 | 50.541 | I | S-1.3m |
| 3425.82024 | 16.904 | -0.163 | 0.027 | 0.769 | 51.252 | 50.541 | I | S-1.3m |
| 3426.75482 | 18.188 | 1.122 | 0.031 | 0.759 | 51.251 | 50.532 | B | S-1.3m |
| 3426.75793 | 17.550 | 0.484 | 0.029 | 0.759 | 51.251 | 50.532 | V | S-1.3m |
| 3426.76003 | 16.830 | -0.236 | 0.044 | 0.759 | 51.251 | 50.532 | I | S-1.3m |
| 3426.76207 | 16.901 | -0.165 | 0.034 | 0.759 | 51.251 | 50.532 | I | S-1.3m |
| 3427.78901 | 18.244 | 1.178 | 0.043 | 0.748 | 51.251 | 50.522 | B | S-1.3m |
| 3427.79212 | 17.799 | 0.733 | 0.045 | 0.748 | 51.251 | 50.522 | V | S-1.3m |
| 3427.79422 | 17.016 | -0.050 | 0.042 | 0.748 | 51.251 | 50.522 | I | S-1.3m |
| 3427.79626 | 17.062 | -0.004 | 0.056 | 0.748 | 51.251 | 50.522 | I | S-1.3m |
| 3428.81384 | 18.075 | 1.009 | 0.034 | 0.738 | 51.251 | 50.514 | B | S-1.3m |
| 3428.81696 | 17.532 | 0.466 | 0.027 | 0.738 | 51.251 | 50.513 | V | S-1.3m |
| 3428.81906 | 16.583 | -0.483 | 0.033 | 0.738 | 51.251 | 50.513 | I | S-1.3m |
| 3428.82109 | 16.741 | -0.325 | 0.034 | 0.737 | 51.251 | 50.513 | I | S-1.3m |
| 3430.81496 | 18.247 | 1.182 | 0.042 | 0.717 | 51.251 | 50.496 | B | S-1.3m |
| 3430.81807 | 17.459 | 0.394 | 0.034 | 0.717 | 51.251 | 50.496 | V | S-1.3m |
| 3430.82018 | 16.916 | -0.149 | 0.032 | 0.717 | 51.251 | 50.496 | I | S-1.3m |
| 3430.82220 | 16.938 | -0.127 | 0.025 | 0.716 | 51.251 | 50.496 | I | S-1.3m |
| 3431.80618 | 18.193 | 1.129 | 0.025 | 0.706 | 51.251 | 50.489 | B | S-1.3m |
| 3431.80929 | 17.602 | 0.538 | 0.028 | 0.706 | 51.251 | 50.489 | V | S-1.3m |
| 3431.81139 | 16.933 | -0.131 | 0.029 | 0.706 | 51.251 | 50.489 | I | S-1.3m |
| 3431.81342 | 16.854 | -0.210 | 0.028 | 0.706 | 51.251 | 50.489 | I | S-1.3m |
| 3432.90631 | 17.559 | 0.495 | 0.025 | 0.695 | 51.251 | 50.480 | V | S-1.3m |
| 3432.90841 | 16.830 | -0.234 | 0.029 | 0.695 | 51.251 | 50.480 | I | S-1.3m |
| 3432.91044 | 16.842 | -0.222 | 0.035 | 0.695 | 51.251 | 50.480 | I | S-1.3m |
| 3433.78165 | 18.185 | 1.121 | 0.018 | 0.685 | 51.251 | 50.473 | B | S-1.3m |
| 3433.78475 | 17.553 | 0.489 | 0.021 | 0.685 | 51.251 | 50.473 | V | S-1.3m |
| 3433.78685 | 16.830 | -0.234 | 0.022 | 0.685 | 51.251 | 50.473 | I | S-1.3m |
| 3433.78887 | 16.789 | -0.275 | 0.024 | 0.685 | 51.251 | 50.473 | I | S-1.3m |
| 3434.76487 | 17.500 | 0.436 | 0.022 | 0.675 | 51.251 | 50.466 | V | S-1.3m |
| 3434.76697 | 16.816 | -0.248 | 0.023 | 0.675 | 51.251 | 50.466 | I | S-1.3m |
| 3434.76899 | 16.814 | -0.250 | 0.022 | 0.675 | 51.251 | 50.466 | I | S-1.3m |
| 3435.81497 | 18.165 | 1.102 | 0.017 | 0.665 | 51.251 | 50.458 | B | S-1.3m |
| 3435.81809 | 17.514 | 0.451 | 0.021 | 0.665 | 51.251 | 50.458 | V | S-1.3m |
| 3435.82019 | 16.832 | -0.231 | 0.021 | 0.665 | 51.251 | 50.458 | I | S-1.3m |
| 3435.82222 | 16.799 | -0.264 | 0.023 | 0.665 | 51.251 | 50.458 | I | S-1.3m |
| 3436.79206 | 18.205 | 1.142 | 0.020 | 0.655 | 51.250 | 50.451 | B | S-1.3m |
| 3436.79516 | 17.548 | 0.485 | 0.024 | 0.655 | 51.250 | 50.451 | V | S-1.3m |
| 3436.79727 | 16.903 | -0.160 | 0.042 | 0.655 | 51.250 | 50.451 | I | S-1.3m |
| 3436.79930 | 16.771 | -0.292 | 0.048 | 0.655 | 51.250 | 50.451 | I | S-1.3m |
| 3439.79654 | 18.396 | 1.334 | 0.019 | 0.626 | 51.250 | 50.432 | B | S-1.3m |
| 3439.79965 | 17.659 | 0.597 | 0.023 | 0.626 | 51.250 | 50.432 | V | S-1.3m |
| 3439.80175 | 16.973 | -0.089 | 0.026 | 0.626 | 51.250 | 50.432 | I | S-1.3m |
| 3439.80378 | 16.911 | -0.151 | 0.023 | 0.626 | 51.250 | 50.432 | I | S-1.3m |
| 3442.77961 | 18.122 | 1.061 | 0.147 | 0.599 | 51.250 | 50.415 | B | S-1.3m |
| 3442.78271 | 17.678 | 0.617 | 0.020 | 0.599 | 51.250 | 50.415 | V | S-1.3m |
| 3442.78481 | 17.042 | -0.019 | 0.021 | 0.599 | 51.250 | 50.415 | I | S-1.3m |
| 3442.78683 | 17.020 | -0.041 | 0.027 | 0.599 | 51.250 | 50.415 | I | S-1.3m |
| 3443.73635 | 18.124 | 1.063 | 0.021 | 0.591 | 51.250 | 50.410 | B | S-1.3m |
| 3443.74154 | 16.766 | -0.295 | 0.021 | 0.591 | 51.250 | 50.410 | I | S-1.3m |
| 3443.74357 | 16.808 | -0.253 | 0.021 | 0.591 | 51.250 | 50.410 | I | S-1.3m |
| 3444.74080 | 18.301 | 1.240 | 0.020 | 0.582 | 51.250 | 50.405 | B | S-1.3m |
| 3444.74391 | 17.672 | 0.611 | 0.021 | 0.582 | 51.250 | 50.405 | V | S-1.3m |
| 3444.74601 | 16.988 | -0.073 | 0.023 | 0.582 | 51.250 | 50.405 | I | S-1.3m |
| 3444.74805 | 17.001 | -0.060 | 0.023 | 0.582 | 51.250 | 50.405 | I | S-1.3m |
| 3445.79097 | 18.113 | 1.052 | 0.017 | 0.574 | 51.249 | 50.400 | B | S-1.3m |
| 3445.79408 | 17.494 | 0.433 | 0.019 | 0.574 | 51.249 | 50.400 | V | S-1.3m |
| 3445.79619 | 16.813 | -0.248 | 0.019 | 0.574 | 51.249 | 50.400 | I | S-1.3m |
| 3445.79822 | 16.844 | -0.217 | 0.021 | 0.574 | 51.249 | 50.400 | I | S-1.3m |
| 3446.74007 | 18.231 | 1.171 | 0.018 | 0.567 | 51.249 | 50.396 | B | S-1.3m |
| 3446.74318 | 17.568 | 0.508 | 0.021 | 0.567 | 51.249 | 50.396 | V | S-1.3m |
| 3446.74528 | 16.818 | -0.242 | 0.021 | 0.567 | 51.249 | 50.396 | I | S-1.3m |
| 3447.72105 | 18.159 | 1.099 | 0.017 | 0.559 | 51.249 | 50.391 | B | S-1.3m |
| 3447.72416 | 17.462 | 0.402 | 0.019 | 0.559 | 51.249 | 50.391 | V | S-1.3m |
| 3447.72627 | 16.773 | -0.287 | 0.019 | 0.559 | 51.249 | 50.391 | I | S-1.3m |
| 3447.72829 | 16.775 | -0.285 | 0.020 | 0.559 | 51.249 | 50.391 | I | S-1.3m |
| 3448.76384 | 18.251 | 1.191 | 0.023 | 0.552 | 51.249 | 50.387 | B | S-1.3m |
| 3448.76694 | 17.580 | 0.520 | 0.027 | 0.552 | 51.249 | 50.387 | V | S-1.3m |
| 3448.76904 | 16.938 | -0.122 | 0.029 | 0.552 | 51.249 | 50.387 | I | S-1.3m |
| 3448.77106 | 16.939 | -0.121 | 0.031 | 0.552 | 51.249 | 50.387 | I | S-1.3m |
| 3449.83314 | 18.198 | 1.138 | 0.027 | 0.546 | 51.249 | 50.384 | B | S-1.3m |
| 3449.83626 | 17.565 | 0.505 | 0.019 | 0.546 | 51.249 | 50.384 | V | S-1.3m |
| 3449.83837 | 16.819 | -0.241 | 0.021 | 0.546 | 51.249 | 50.384 | I | S-1.3m |
| 3449.84039 | 16.824 | -0.236 | 0.021 | 0.546 | 51.249 | 50.384 | I | S-1.3m |
| 3450.77092 | 18.172 | 1.112 | 0.017 | 0.540 | 51.249 | 50.380 | B | S-1.3m |
| 3450.77402 | 17.542 | 0.482 | 0.019 | 0.540 | 51.249 | 50.380 | V | S-1.3m |
| 3450.77612 | 16.909 | -0.151 | 0.020 | 0.540 | 51.249 | 50.380 | I | S-1.3m |
| 3450.77814 | 16.900 | -0.160 | 0.020 | 0.540 | 51.249 | 50.380 | I | S-1.3m |
| 3453.76541 | 18.097 | 1.038 | 0.023 | 0.525 | 51.249 | 50.371 | B | S-1.3m |
| 3453.76845 | 17.447 | 0.388 | 0.024 | 0.525 | 51.249 | 50.371 | V | S-1.3m |
| 3453.77047 | 16.726 | -0.333 | 0.023 | 0.525 | 51.249 | 50.371 | I | S-1.3m |
| 3453.77244 | 16.767 | -0.292 | 0.023 | 0.525 | 51.249 | 50.371 | I | S-1.3m |
| 3458.76705 | 18.197 | 1.138 | 0.033 | 0.510 | 51.248 | 50.363 | B | S-1.3m |
| 3458.77005 | 17.583 | 0.524 | 0.027 | 0.510 | 51.248 | 50.363 | V | S-1.3m |
| 3458.77204 | 16.911 | -0.148 | 0.035 | 0.510 | 51.248 | 50.363 | I | S-1.3m |
| 3458.77398 | 16.880 | -0.179 | 0.040 | 0.510 | 51.248 | 50.363 | I | S-1.3m |
| 3459.77931 | 18.246 | 1.187 | 0.026 | 0.509 | 51.248 | 50.362 | B | S-1.3m |
| 3459.78236 | 17.647 | 0.588 | 0.029 | 0.509 | 51.248 | 50.362 | V | S-1.3m |
| 3459.78439 | 16.948 | -0.111 | 0.040 | 0.509 | 51.248 | 50.362 | I | S-1.3m |
| 3459.78636 | 16.971 | -0.088 | 0.037 | 0.509 | 51.248 | 50.362 | I | S-1.3m |
| 3460.73187 | 18.332 | 1.273 | 0.023 | 0.508 | 51.248 | 50.361 | B | S-1.3m |
| 3460.73490 | 17.715 | 0.656 | 0.023 | 0.508 | 51.248 | 50.361 | V | S-1.3m |
| 3460.73889 | 17.061 | 0.002 | 0.024 | 0.508 | 51.248 | 50.361 | I | S-1.3m |
| 3461.75094 | 18.229 | 1.170 | 0.021 | 0.508 | 51.248 | 50.361 | B | S-1.3m |
| 3461.75399 | 17.537 | 0.478 | 0.024 | 0.508 | 51.248 | 50.361 | V | S-1.3m |
| 3461.75602 | 16.847 | -0.212 | 0.023 | 0.508 | 51.248 | 50.361 | I | S-1.3m |
| 3461.75798 | 16.825 | -0.234 | 0.023 | 0.508 | 51.248 | 50.361 | I | S-1.3m |
| 3462.70513 | 18.363 | 1.304 | 0.019 | 0.509 | 51.248 | 50.360 | B | S-1.3m |
| 3462.70818 | 17.729 | 0.670 | 0.021 | 0.509 | 51.248 | 50.360 | V | S-1.3m |
| 3462.71021 | 17.002 | -0.057 | 0.021 | 0.509 | 51.248 | 50.360 | I | S-1.3m |
| 3462.71217 | 17.004 | -0.055 | 0.021 | 0.509 | 51.248 | 50.360 | I | S-1.3m |
| 3463.73429 | 18.061 | 1.002 | 0.018 | 0.510 | 51.248 | 50.360 | B | S-1.3m |

| | | | | | | | | | |
|---|---|---|---|---|---|---|---|---|---|
| 3463.73732 | 17.444 | 0.385 | 0.021 | 0.510 | 51.248 | 50.360 | V | S-1.3m | |
| 3463.73935 | 16.817 | -0.242 | 0.019 | 0.510 | 51.248 | 50.360 | I | S-1.3m | |
| 3463.74132 | 16.822 | -0.237 | 0.021 | 0.510 | 51.248 | 50.360 | I | S-1.3m | |
| 3464.69544 | 18.110 | 1.051 | 0.019 | 0.511 | 51.247 | 50.361 | B | S-1.3m | |
| 3464.69847 | 17.432 | 0.373 | 0.019 | 0.511 | 51.247 | 50.361 | V | S-1.3m | |
| 3464.70050 | 16.750 | -0.309 | 0.021 | 0.511 | 51.247 | 50.361 | I | S-1.3m | |
| 3464.70246 | 16.697 | -0.362 | 0.023 | 0.511 | 51.247 | 50.361 | I | S-1.3m | |
| 3465.74591 | 18.192 | 1.133 | 0.018 | 0.514 | 51.247 | 50.361 | B | S-1.3m | |
| 3465.74894 | 17.520 | 0.461 | 0.020 | 0.514 | 51.247 | 50.361 | V | S-1.3m | |
| 3465.75097 | 16.846 | -0.213 | 0.020 | 0.514 | 51.247 | 50.361 | I | S-1.3m | |
| 3465.75294 | 16.794 | -0.265 | 0.021 | 0.514 | 51.247 | 50.361 | I | S-1.3m | |
| 3466.70574 | 18.296 | 1.237 | 0.019 | 0.517 | 51.247 | 50.362 | B | S-1.3m | |
| 3466.70877 | 17.647 | 0.588 | 0.021 | 0.517 | 51.247 | 50.362 | V | S-1.3m | |
| 3466.71080 | 16.977 | -0.082 | 0.023 | 0.517 | 51.247 | 50.362 | I | S-1.3m | |
| 3466.71276 | 16.984 | -0.075 | 0.023 | 0.517 | 51.247 | 50.362 | I | S-1.3m | |
| 3467.72113 | 18.039 | 0.980 | 0.019 | 0.520 | 51.247 | 50.363 | B | S-1.3m | |
| 3467.72416 | 17.433 | 0.374 | 0.022 | 0.520 | 51.247 | 50.363 | V | S-1.3m | |
| 3467.72620 | 16.712 | -0.347 | 0.021 | 0.520 | 51.247 | 50.363 | I | S-1.3m | |
| 3467.72816 | 16.733 | -0.326 | 0.023 | 0.520 | 51.247 | 50.363 | I | S-1.3m | |
| 3468.68440 | 17.778 | 0.719 | 0.023 | 0.524 | 51.247 | 50.364 | B | S-1.3m | rejected |
| 3468.68746 | 17.447 | 0.388 | 0.023 | 0.524 | 51.247 | 50.364 | V | S-1.3m | |
| 3468.68949 | 16.803 | -0.256 | 0.034 | 0.524 | 51.247 | 50.364 | I | S-1.3m | |
| 3468.69146 | 16.789 | -0.270 | 0.027 | 0.524 | 51.247 | 50.364 | I | S-1.3m | |
| 3469.62949 | 18.215 | 1.156 | 0.019 | 0.529 | 51.247 | 50.366 | B | S-1.3m | |
| 3469.63252 | 17.614 | 0.555 | 0.023 | 0.529 | 51.247 | 50.366 | V | S-1.3m | |
| 3469.63456 | 16.939 | -0.120 | 0.021 | 0.529 | 51.247 | 50.366 | I | S-1.3m | |
| 3469.63652 | 16.952 | -0.107 | 0.023 | 0.529 | 51.247 | 50.366 | I | S-1.3m | |
| 3470.71393 | 18.250 | 1.191 | 0.018 | 0.534 | 51.247 | 50.368 | B | S-1.3m | |
| 3470.71696 | 17.567 | 0.508 | 0.019 | 0.534 | 51.247 | 50.368 | V | S-1.3m | |
| 3470.71899 | 16.854 | -0.205 | 0.021 | 0.534 | 51.247 | 50.368 | I | S-1.3m | |
| 3470.72096 | 16.832 | -0.227 | 0.021 | 0.534 | 51.247 | 50.368 | I | S-1.3m | |
| 3471.79822 | 17.075 | 0.039 | 0.050 | 0.540 | 51.247 | 50.371 | r | P-200 | relative |
| 3471.80028 | 17.060 | 0.024 | 0.050 | 0.540 | 51.247 | 50.371 | r | P-200 | relative |
| 3471.80236 | 17.092 | 0.056 | 0.050 | 0.540 | 51.247 | 50.371 | r | P-200 | relative |
| 3471.80441 | 17.088 | 0.052 | 0.050 | 0.540 | 51.247 | 50.371 | r | P-200 | relative |
| 3471.84993 | 17.271 | 0.235 | 0.050 | 0.540 | 51.247 | 50.371 | r | P-200 | relative |
| 3471.87415 | 17.052 | 0.016 | 0.050 | 0.540 | 51.247 | 50.371 | r | P-200 | relative |
| 3471.89804 | 17.088 | 0.052 | 0.050 | 0.540 | 51.247 | 50.371 | r | P-200 | relative |
| 3471.91587 | 17.242 | 0.206 | 0.050 | 0.540 | 51.247 | 50.371 | r | P-200 | relative |
| 3471.93398 | 17.266 | 0.230 | 0.050 | 0.541 | 51.247 | 50.371 | r | P-200 | relative |
| 3471.94970 | 17.142 | 0.106 | 0.050 | 0.541 | 51.247 | 50.371 | r | P-200 | relative |
| 3471.96962 | 17.127 | 0.091 | 0.050 | 0.541 | 51.247 | 50.371 | r | P-200 | relative |
| 3471.98803 | 17.285 | 0.249 | 0.050 | 0.541 | 51.247 | 50.371 | r | P-200 | relative |
| 3472.00370 | 17.350 | 0.314 | 0.050 | 0.541 | 51.247 | 50.372 | r | P-200 | relative |
| 3472.00576 | 17.324 | 0.288 | 0.050 | 0.541 | 51.247 | 50.372 | r | P-200 | relative |
| 3472.00780 | 17.345 | 0.309 | 0.050 | 0.541 | 51.247 | 50.372 | r | P-200 | relative |
| 3472.00985 | 17.333 | 0.297 | 0.050 | 0.541 | 51.247 | 50.372 | r | P-200 | relative |
| 3472.01190 | 17.330 | 0.294 | 0.050 | 0.541 | 51.247 | 50.372 | r | P-200 | relative |
| 3472.01394 | 17.298 | 0.262 | 0.050 | 0.541 | 51.247 | 50.372 | r | P-200 | relative |
| 3472.01599 | 17.308 | 0.272 | 0.050 | 0.541 | 51.247 | 50.372 | r | P-200 | relative |
| 3472.01805 | 17.275 | 0.239 | 0.050 | 0.541 | 51.247 | 50.372 | r | P-200 | relative |
| 3472.02010 | 17.263 | 0.227 | 0.050 | 0.541 | 51.247 | 50.372 | r | P-200 | relative |
| 3472.71209 | 17.110 | 0.074 | 0.050 | 0.545 | 51.247 | 50.373 | r | P-200 | relative |
| 3472.73583 | 17.291 | 0.255 | 0.050 | 0.545 | 51.247 | 50.373 | r | P-200 | relative |
| 3472.75953 | 17.215 | 0.179 | 0.050 | 0.546 | 51.247 | 50.374 | r | P-200 | relative |
| 3472.78323 | 17.115 | 0.079 | 0.050 | 0.546 | 51.247 | 50.374 | r | P-200 | relative |
| 3472.80681 | 17.295 | 0.259 | 0.050 | 0.546 | 51.247 | 50.374 | r | P-200 | relative |
| 3472.85269 | 17.069 | 0.033 | 0.050 | 0.546 | 51.247 | 50.374 | r | P-200 | relative |
| 3472.96031 | 17.216 | 0.180 | 0.050 | 0.547 | 51.247 | 50.374 | r | P-200 | relative |
| 3473.00560 | 17.289 | 0.253 | 0.050 | 0.547 | 51.247 | 50.374 | r | P-200 | relative |
| 3473.00763 | 17.274 | 0.238 | 0.050 | 0.547 | 51.247 | 50.374 | r | P-200 | relative |
| 3473.00984 | 17.203 | 0.167 | 0.050 | 0.547 | 51.247 | 50.374 | r | P-200 | relative |
| 3473.01189 | 17.231 | 0.195 | 0.050 | 0.547 | 51.247 | 50.374 | r | P-200 | relative |
| 3473.01393 | 17.202 | 0.166 | 0.050 | 0.547 | 51.247 | 50.374 | r | P-200 | relative |
| 3473.01601 | 17.194 | 0.158 | 0.050 | 0.547 | 51.247 | 50.374 | r | P-200 | relative |
| 3473.01805 | 17.174 | 0.138 | 0.050 | 0.547 | 51.247 | 50.374 | r | P-200 | relative |
| 3473.02010 | 17.176 | 0.140 | 0.050 | 0.547 | 51.247 | 50.374 | r | P-200 | relative |
| 3473.75572 | 18.095 | 1.036 | 0.023 | 0.552 | 51.247 | 50.376 | B | S-1.3m | |
| 3473.75876 | 17.437 | 0.378 | 0.024 | 0.552 | 51.247 | 50.376 | V | S-1.3m | |
| 3473.76079 | 16.734 | -0.325 | 0.027 | 0.552 | 51.247 | 50.376 | I | S-1.3m | |
| 3473.76275 | 16.777 | -0.282 | 0.024 | 0.552 | 51.247 | 50.376 | I | S-1.3m | |
| 3474.76664 | 18.301 | 1.241 | 0.019 | 0.559 | 51.246 | 50.379 | B | S-1.3m | |
| 3474.76970 | 17.662 | 0.602 | 0.022 | 0.559 | 51.246 | 50.379 | V | S-1.3m | |
| 3474.77172 | 16.996 | -0.064 | 0.024 | 0.559 | 51.246 | 50.379 | I | S-1.3m | |
| 3474.77369 | 17.015 | -0.045 | 0.023 | 0.559 | 51.246 | 50.379 | I | S-1.3m | |
| 3475.74948 | 18.305 | 1.245 | 0.018 | 0.566 | 51.246 | 50.382 | B | S-1.3m | |
| 3475.75252 | 17.638 | 0.578 | 0.021 | 0.566 | 51.246 | 50.382 | V | S-1.3m | |
| 3475.75454 | 16.925 | -0.135 | 0.030 | 0.566 | 51.246 | 50.382 | I | S-1.3m | |
| 3476.76649 | 18.081 | 1.021 | 0.454 | 0.573 | 51.246 | 50.386 | B | S-1.3m | rejected |
| 3476.76954 | 17.360 | 0.300 | 0.676 | 0.573 | 51.246 | 50.386 | V | S-1.3m | rejected |
| 3476.77157 | 16.673 | -0.387 | 0.558 | 0.573 | 51.246 | 50.386 | I | S-1.3m | rejected |
| 3476.77354 | 16.655 | -0.405 | 0.577 | 0.573 | 51.246 | 50.386 | I | S-1.3m | rejected |
| 3477.72766 | 18.216 | 1.156 | 0.018 | 0.581 | 51.246 | 50.389 | B | S-1.3m | |
| 3477.73071 | 17.456 | 0.396 | 0.028 | 0.581 | 51.246 | 50.389 | V | S-1.3m | |
| 3477.73274 | 16.622 | -0.438 | 0.043 | 0.581 | 51.246 | 50.389 | I | S-1.3m | |
| 3477.73471 | 16.551 | -0.509 | 0.039 | 0.581 | 51.246 | 50.389 | I | S-1.3m | rejected |
| 3478.66834 | 18.084 | 1.024 | 0.048 | 0.589 | 51.246 | 50.393 | B | S-1.3m | |
| 3478.67140 | 17.444 | 0.384 | 0.047 | 0.589 | 51.246 | 50.393 | V | S-1.3m | |
| 3478.67343 | 16.632 | -0.428 | 0.042 | 0.589 | 51.246 | 50.393 | I | S-1.3m | |
| 3478.67540 | 16.668 | -0.392 | 0.032 | 0.589 | 51.246 | 50.393 | I | S-1.3m | |
| 3486.64441 | 18.035 | 0.973 | 0.060 | 0.664 | 51.245 | 50.434 | B | S-1.3m | |
| 3486.64745 | 17.315 | 0.253 | 0.075 | 0.664 | 51.245 | 50.434 | V | S-1.3m | |
| 3486.64948 | 16.872 | -0.190 | 0.072 | 0.664 | 51.245 | 50.434 | I | S-1.3m | |
| 3486.65145 | 16.862 | -0.200 | 0.041 | 0.664 | 51.245 | 50.434 | I | S-1.3m | |
| 3489.68064 | 18.111 | 1.048 | 0.248 | 0.695 | 51.245 | 50.453 | B | S-1.3m | |
| 3489.68370 | 17.313 | 0.250 | 0.052 | 0.695 | 51.245 | 50.453 | V | S-1.3m | |
| 3489.68769 | 16.886 | -0.177 | 0.036 | 0.695 | 51.245 | 50.453 | I | S-1.3m | |
| 3490.69451 | 18.343 | 1.280 | 0.028 | 0.706 | 51.245 | 50.460 | B | S-1.3m | |
| 3490.69757 | 17.575 | 0.512 | 0.033 | 0.706 | 51.245 | 50.460 | V | S-1.3m | |
| 3490.69960 | 16.978 | -0.085 | 0.025 | 0.706 | 51.245 | 50.460 | I | S-1.3m | |
| 3490.70156 | 16.978 | -0.085 | 0.039 | 0.706 | 51.245 | 50.460 | I | S-1.3m | |
| 3491.70627 | 18.160 | 1.097 | 0.024 | 0.716 | 51.245 | 50.468 | B | S-1.3m | |
| 3491.70933 | 17.399 | 0.336 | 0.026 | 0.716 | 51.245 | 50.468 | V | S-1.3m | |
| 3491.71136 | 16.855 | -0.208 | 0.026 | 0.716 | 51.245 | 50.468 | I | S-1.3m | |
| 3491.71332 | 16.859 | -0.204 | 0.024 | 0.716 | 51.245 | 50.468 | I | S-1.3m | |
| 3492.61333 | 18.097 | 1.033 | 0.034 | 0.726 | 51.244 | 50.475 | B | S-1.3m | |
| 3492.61637 | 17.500 | 0.436 | 0.041 | 0.726 | 51.244 | 50.475 | V | S-1.3m | |
| 3492.61841 | 16.791 | -0.273 | 0.056 | 0.726 | 51.244 | 50.475 | I | S-1.3m | |
| 3492.62038 | 16.891 | -0.173 | 0.027 | 0.726 | 51.244 | 50.475 | I | S-1.3m | |
| 3494.66097 | 0.043 | 0.061 | 0.100 | 0.748 | 51.244 | 50.490 | R | T-32 | relative |
| 3494.68157 | 0.140 | 0.158 | 0.100 | 0.748 | 51.244 | 50.490 | R | T-32 | relative |
| 3494.68907 | 0.082 | 0.100 | 0.100 | 0.748 | 51.244 | 50.490 | R | T-32 | relative |
| 3494.70954 | -0.113 | -0.095 | 0.100 | 0.748 | 51.244 | 50.491 | R | T-32 | relative |
| 3494.71659 | -0.148 | -0.130 | 0.100 | 0.748 | 51.244 | 50.491 | R | T-32 | relative |
| 3494.72328 | -0.236 | -0.218 | 0.100 | 0.748 | 51.244 | 50.491 | R | T-32 | relative |
| 3494.72994 | -0.194 | -0.176 | 0.100 | 0.749 | 51.244 | 50.491 | R | T-32 | relative |
| 3494.74476 | -0.087 | -0.069 | 0.100 | 0.749 | 51.244 | 50.491 | R | T-32 | relative |
| 3494.76453 | 0.052 | 0.070 | 0.100 | 0.749 | 51.244 | 50.491 | R | T-32 | relative |
| 3494.77142 | 0.136 | 0.154 | 0.100 | 0.749 | 51.244 | 50.491 | R | T-32 | relative |
| 3494.77835 | 0.040 | 0.058 | 0.100 | 0.749 | 51.244 | 50.491 | R | T-32 | relative |
| 3494.78575 | 0.008 | 0.026 | 0.100 | 0.749 | 51.244 | 50.491 | R | T-32 | relative |
| 3494.79271 | -0.111 | -0.093 | 0.100 | 0.749 | 51.244 | 50.491 | R | T-32 | relative |
| 3494.79955 | -0.187 | -0.169 | 0.100 | 0.749 | 51.244 | 50.491 | R | T-32 | relative |
| 3494.81366 | -0.159 | -0.141 | 0.100 | 0.749 | 51.244 | 50.491 | R | T-32 | relative |
| 3494.82055 | -0.055 | -0.037 | 0.100 | 0.750 | 51.244 | 50.492 | R | T-32 | relative |
| 3494.82970 | 0.117 | 0.135 | 0.100 | 0.750 | 51.244 | 50.492 | R | T-32 | relative |
| 3494.83681 | 0.129 | 0.147 | 0.100 | 0.750 | 51.244 | 50.492 | R | T-32 | relative |
| 3494.84363 | 0.201 | 0.219 | 0.100 | 0.750 | 51.244 | 50.492 | R | T-32 | relative |
| 3494.85069 | 0.000 | 0.018 | 0.100 | 0.750 | 51.244 | 50.492 | R | T-32 | relative |
| 3494.85774 | 0.012 | 0.030 | 0.100 | 0.750 | 51.244 | 50.492 | R | T-32 | relative |
| 3494.86477 | -0.142 | -0.124 | 0.100 | 0.750 | 51.244 | 50.492 | R | T-32 | relative |
| 3504.61968 | 18.303 | 1.235 | 0.021 | 0.853 | 51.243 | 50.580 | B | S-1.3m | |
| 3504.62270 | 17.633 | 0.565 | 0.026 | 0.853 | 51.243 | 50.580 | V | S-1.3m | |
| 3504.62669 | 16.986 | -0.082 | 0.024 | 0.853 | 51.243 | 50.580 | I | S-1.3m | |
| 3507.57248 | 18.309 | 1.240 | 0.063 | 0.884 | 51.243 | 50.610 | B | S-1.3m | |
| 3507.57552 | 17.652 | 0.583 | 0.193 | 0.884 | 51.243 | 50.610 | V | S-1.3m | |
| 3507.57755 | 16.946 | -0.123 | 0.578 | 0.884 | 51.243 | 50.610 | I | S-1.3m | rejected |
| 3507.57951 | 17.191 | 0.122 | 0.701 | 0.884 | 51.243 | 50.610 | I | S-1.3m | rejected |
| 3508.56663 | 18.286 | 1.216 | 0.024 | 0.894 | 51.243 | 50.621 | B | S-1.3m | |
| 3508.56968 | 17.623 | 0.553 | 0.026 | 0.894 | 51.243 | 50.621 | V | S-1.3m | |
| 3508.57170 | 16.898 | -0.172 | 0.026 | 0.894 | 51.243 | 50.621 | I | S-1.3m | |
| 3509.64165 | 18.165 | 1.095 | 0.024 | 0.903 | 51.243 | 50.633 | B | S-1.3m | |
| 3509.64469 | 17.500 | 0.430 | 0.025 | 0.903 | 51.243 | 50.633 | V | S-1.3m | |
| 3509.64672 | 16.807 | -0.263 | 0.026 | 0.903 | 51.243 | 50.633 | I | S-1.3m | |
| 3509.64869 | 16.816 | -0.254 | 0.027 | 0.903 | 51.243 | 50.633 | I | S-1.3m | |
| 3515.61182 | 16.817 | -0.256 | 0.032 | 0.959 | 51.242 | 50.700 | I | S-1.3m | |
| 3515.61379 | 16.743 | -0.330 | 0.029 | 0.959 | 51.242 | 50.700 | I | S-1.3m | |

| | | | | | | | | |
|---|---|---|---|---|---|---|---|---|
| 3517.59339 | 18.332 | 1.258 | 0.025 | 0.976 | 51.242 | 50.724 | B | S-1.3m |
| 3517.59643 | 17.643 | 0.569 | 0.024 | 0.976 | 51.242 | 50.724 | V | S-1.3m |
| 3517.59847 | 16.995 | -0.079 | 0.025 | 0.976 | 51.242 | 50.724 | I | S-1.3m |
| 3517.60043 | 17.002 | -0.072 | 0.025 | 0.976 | 51.242 | 50.724 | I | S-1.3m |
| 3518.64243 | 18.312 | 1.237 | 0.035 | 0.984 | 51.242 | 50.738 | B | S-1.3m |
| 3518.64547 | 17.646 | 0.571 | 0.032 | 0.984 | 51.242 | 50.738 | V | S-1.3m |
| 3518.64750 | 17.029 | -0.046 | 0.024 | 0.984 | 51.242 | 50.738 | I | S-1.3m |
| 3518.64947 | 17.036 | -0.039 | 0.025 | 0.984 | 51.242 | 50.738 | I | S-1.3m |
| 3520.61043 | 18.403 | 1.327 | 0.019 | 1.001 | 51.241 | 50.761 | B | S-1.3m |
| 3520.61347 | 17.703 | 0.627 | 0.021 | 1.001 | 51.241 | 50.761 | V | S-1.3m |
| 3520.61549 | 17.056 | -0.020 | 0.024 | 1.001 | 51.241 | 50.761 | I | S-1.3m |
| 3520.61746 | 17.081 | 0.005 | 0.025 | 1.001 | 51.241 | 50.762 | I | S-1.3m |
| 3521.59396 | 18.429 | 1.353 | 0.021 | 1.008 | 51.241 | 50.774 | B | S-1.3m |
| 3521.59700 | 17.769 | 0.693 | 0.024 | 1.008 | 51.241 | 50.774 | V | S-1.3m |
| 3521.59903 | 17.108 | 0.032 | 0.028 | 1.008 | 51.241 | 50.774 | I | S-1.3m |
| 3521.60100 | 17.078 | 0.002 | 0.027 | 1.008 | 51.241 | 50.774 | I | S-1.3m |
| 3522.59185 | 18.355 | 1.278 | 0.017 | 1.016 | 51.241 | 50.787 | B | S-1.3m |
| 3522.59488 | 17.676 | 0.599 | 0.020 | 1.016 | 51.241 | 50.787 | V | S-1.3m |
| 3522.59691 | 16.987 | -0.090 | 0.023 | 1.016 | 51.241 | 50.787 | I | S-1.3m |
| 3522.59888 | 16.932 | -0.145 | 0.021 | 1.016 | 51.241 | 50.787 | I | S-1.3m |
| 3523.56347 | 18.365 | 1.288 | 0.019 | 1.023 | 51.241 | 50.800 | B | S-1.3m |
| 3523.56652 | 17.721 | 0.644 | 0.022 | 1.023 | 51.241 | 50.800 | V | S-1.3m |
| 3523.56855 | 17.043 | -0.034 | 0.021 | 1.023 | 51.241 | 50.800 | I | S-1.3m |
| 3523.57052 | 16.992 | -0.085 | 0.023 | 1.023 | 51.241 | 50.800 | I | S-1.3m |
| 3524.55372 | 18.247 | 1.169 | 0.022 | 1.030 | 51.241 | 50.813 | B | S-1.3m |
| 3524.55677 | 17.628 | 0.550 | 0.032 | 1.030 | 51.241 | 50.813 | V | S-1.3m |
| 3524.55880 | 16.895 | -0.183 | 0.039 | 1.030 | 51.241 | 50.813 | I | S-1.3m |
| 3524.56076 | 16.839 | -0.239 | 0.039 | 1.030 | 51.241 | 50.813 | I | S-1.3m |
| 3525.60217 | 18.325 | 1.246 | 0.102 | 1.038 | 51.241 | 50.826 | B | S-1.3m |
| 3525.60523 | 17.742 | 0.663 | 0.105 | 1.038 | 51.241 | 50.826 | V | S-1.3m |
| 3525.60727 | 16.969 | -0.110 | 0.100 | 1.038 | 51.241 | 50.826 | I | S-1.3m |
| 3525.60924 | 16.958 | -0.121 | 0.092 | 1.038 | 51.241 | 50.827 | I | S-1.3m |
| 3526.59391 | 18.332 | 1.253 | 0.048 | 1.044 | 51.241 | 50.840 | B | S-1.3m |
| 3526.59695 | 17.570 | 0.491 | 0.091 | 1.044 | 51.241 | 50.840 | V | S-1.3m |
| 3526.59898 | 16.872 | -0.207 | 0.070 | 1.044 | 51.241 | 50.840 | I | S-1.3m |
| 3526.60095 | 16.964 | -0.115 | 0.114 | 1.044 | 51.241 | 50.840 | I | S-1.3m |
| 3527.54428 | 18.270 | 1.190 | 0.028 | 1.050 | 51.241 | 50.853 | B | S-1.3m |
| 3527.54733 | 17.616 | 0.536 | 0.052 | 1.050 | 51.241 | 50.853 | V | S-1.3m |
| 3527.54936 | 17.043 | -0.037 | 0.062 | 1.050 | 51.241 | 50.853 | I | S-1.3m |
| 3527.55133 | 16.948 | -0.132 | 0.038 | 1.050 | 51.241 | 50.853 | I | S-1.3m |
| 3528.57725 | 18.198 | 1.118 | 0.019 | 1.056 | 51.241 | 50.867 | B | S-1.3m |
| 3528.58028 | 17.578 | 0.498 | 0.021 | 1.056 | 51.241 | 50.867 | V | S-1.3m |
| 3528.58230 | 16.881 | -0.199 | 0.029 | 1.056 | 51.241 | 50.867 | I | S-1.3m |
| 3528.58426 | 16.914 | -0.166 | 0.038 | 1.057 | 51.241 | 50.867 | I | S-1.3m |
| 3529.54537 | 18.298 | 1.217 | 0.096 | 1.062 | 51.241 | 50.880 | B | S-1.3m |
| 3529.54841 | 17.723 | 0.642 | 0.177 | 1.062 | 51.241 | 50.880 | V | S-1.3m |
| 3529.55044 | 16.861 | -0.220 | 0.032 | 1.062 | 51.241 | 50.880 | I | S-1.3m |
| 3529.55240 | 16.850 | -0.231 | 0.037 | 1.062 | 51.241 | 50.880 | I | S-1.3m |
| 3530.50309 | 18.368 | 1.287 | 0.020 | 1.068 | 51.240 | 50.893 | B | S-1.3m |
| 3530.50612 | 17.736 | 0.655 | 0.025 | 1.068 | 51.240 | 50.893 | V | S-1.3m |
| 3530.50815 | 17.034 | -0.047 | 0.029 | 1.068 | 51.240 | 50.893 | I | S-1.3m |
| 3530.51011 | 16.996 | -0.085 | 0.034 | 1.068 | 51.240 | 50.893 | I | S-1.3m |
| 3538.57824 | 17.142 | 0.056 | 0.027 | 1.107 | 51.240 | 51.007 | I | S-1.3m |
| 3538.58019 | 17.179 | 0.093 | 0.029 | 1.107 | 51.240 | 51.007 | I | S-1.3m |
| 3542.54940 | 17.040 | -0.049 | 0.062 | 1.120 | 51.239 | 51.065 | I | S-1.3m |
| 3542.55137 | 17.072 | -0.017 | 0.046 | 1.120 | 51.239 | 51.065 | I | S-1.3m |
| 3543.52344 | 16.952 | -0.137 | 0.038 | 1.123 | 51.239 | 51.079 | I | S-1.3m |
| 3543.52540 | 17.018 | -0.071 | 0.035 | 1.123 | 51.239 | 51.079 | I | S-1.3m |
| 3544.51805 | 17.067 | -0.023 | 0.026 | 1.126 | 51.239 | 51.094 | I | S-1.3m |
| 3544.52002 | 17.120 | 0.030 | 0.024 | 1.126 | 51.239 | 51.094 | I | S-1.3m |
| 3547.44257 | 16.998 | -0.094 | 0.062 | 1.132 | 51.239 | 51.137 | I | S-1.3m |
| 3547.46362 | 17.056 | -0.036 | 0.028 | 1.132 | 51.239 | 51.138 | I | S-1.3m |
| 3547.48131 | 17.048 | -0.044 | 0.038 | 1.132 | 51.239 | 51.138 | I | S-1.3m |
| 3547.50315 | 16.933 | -0.159 | 0.021 | 1.132 | 51.239 | 51.138 | I | S-1.3m |
| 3547.53836 | 17.092 | 0.000 | 0.027 | 1.132 | 51.239 | 51.139 | I | S-1.3m |
| 3547.57037 | 16.816 | -0.276 | 0.027 | 1.132 | 51.239 | 51.139 | I | S-1.3m |
| 3547.58859 | 16.947 | -0.145 | 0.062 | 1.132 | 51.239 | 51.139 | I | S-1.3m |
| 3550.46438 | 17.172 | 0.078 | 0.063 | 1.135 | 51.238 | 51.182 | I | S-1.3m |
| 3550.48285 | 17.235 | 0.141 | 0.057 | 1.135 | 51.238 | 51.183 | I | S-1.3m |
| 3550.50585 | 16.856 | -0.238 | 0.043 | 1.135 | 51.238 | 51.183 | I | S-1.3m |
| 3550.54054 | 17.124 | 0.030 | 0.040 | 1.135 | 51.238 | 51.184 | I | S-1.3m |
| 3550.55194 | 17.037 | -0.057 | 0.043 | 1.135 | 51.238 | 51.184 | I | S-1.3m |
| 3550.58016 | 17.017 | -0.077 | 0.059 | 1.135 | 51.238 | 51.184 | I | S-1.3m |
| 3569.46736 | 18.373 | 1.268 | 0.026 | 1.106 | 51.236 | 51.465 | B | S-1.3m |
| 3569.47040 | 17.749 | 0.644 | 0.026 | 1.106 | 51.236 | 51.465 | V | S-1.3m |
| 3569.47243 | 17.073 | -0.032 | 0.039 | 1.106 | 51.236 | 51.465 | I | S-1.3m |
| 3569.47439 | 17.052 | -0.053 | 0.034 | 1.106 | 51.236 | 51.465 | I | S-1.3m |
| 3570.50356 | 18.164 | 1.058 | 0.035 | 1.102 | 51.236 | 51.479 | B | S-1.3m |
| 3570.50661 | 17.656 | 0.550 | 0.050 | 1.102 | 51.236 | 51.479 | V | S-1.3m |
| 3570.50864 | 16.867 | -0.239 | 0.035 | 1.102 | 51.236 | 51.479 | I | S-1.3m |
| 3570.51061 | 16.866 | -0.240 | 0.037 | 1.102 | 51.236 | 51.479 | I | S-1.3m |
| 3571.50977 | 18.305 | 1.198 | 0.088 | 1.098 | 51.236 | 51.494 | B | S-1.3m |
| 3571.51283 | 17.770 | 0.663 | 0.274 | 1.098 | 51.236 | 51.494 | V | S-1.3m |
| 3571.51486 | 17.073 | -0.034 | 0.139 | 1.098 | 51.236 | 51.494 | I | S-1.3m |
| 3571.51683 | 17.224 | 0.117 | 0.162 | 1.098 | 51.236 | 51.494 | I | S-1.3m |
| 3573.50741 | 17.474 | 0.366 | 0.036 | 1.088 | 51.236 | 51.522 | V | S-1.3m |
| 3573.50846 | 18.177 | 1.069 | 0.046 | 1.088 | 51.236 | 51.522 | B | S-1.3m |
| 3573.51754 | 16.781 | -0.327 | 0.039 | 1.088 | 51.236 | 51.522 | I | S-1.3m |
| 3573.51951 | 16.834 | -0.274 | 0.037 | 1.088 | 51.236 | 51.522 | I | S-1.3m |
| 3574.47016 | 18.352 | 1.244 | 0.020 | 1.083 | 51.236 | 51.536 | B | S-1.3m |
| 3574.47322 | 17.672 | 0.564 | 0.023 | 1.083 | 51.236 | 51.536 | V | S-1.3m |
| 3574.47525 | 16.918 | -0.190 | 0.036 | 1.083 | 51.236 | 51.536 | I | S-1.3m |
| 3574.47722 | 16.974 | -0.134 | 0.034 | 1.083 | 51.236 | 51.536 | I | S-1.3m |
| 3575.48804 | 18.181 | 1.072 | 0.018 | 1.078 | 51.236 | 51.550 | B | S-1.3m |
| 3575.49109 | 17.546 | 0.437 | 0.021 | 1.078 | 51.236 | 51.550 | V | S-1.3m |
| 3575.49312 | 16.876 | -0.233 | 0.024 | 1.078 | 51.236 | 51.550 | I | S-1.3m |
| 3575.86914 | 18.168 | 1.059 | 0.027 | 1.076 | 51.236 | 51.555 | B | S-1.3m |
| 3576.48094 | 18.319 | 1.209 | 0.019 | 1.073 | 51.235 | 51.564 | B | S-1.3m |
| 3576.48400 | 17.701 | 0.591 | 0.023 | 1.073 | 51.235 | 51.564 | V | S-1.3m |
| 3576.48603 | 16.990 | -0.120 | 0.027 | 1.073 | 51.235 | 51.564 | I | S-1.3m |
| 3576.48800 | 16.983 | -0.127 | 0.027 | 1.073 | 51.235 | 51.564 | I | S-1.3m |

Note: The telescopes listed are the SMARTS 1.3m (S-1.3m), the Palomar 200" (P-200) and the Tenagra 32" (T-32).

Table 2. Linear fits to the solar phase curve, reduced magnitude = a * α + b.  in B, V, and I.

| Filter | a (mag deg$^{-1}$) | b (mag) |
|---|---|---|
| B | 0.085 ± 0.020 | 1.081 ± 0.016 |
| V | 0.091 ± 0.025 | 0.444 ± 0.021 |
| I | 0.132 ± 0.033 | -0.259 ± 0.028 |

Table 3. Average reduced magnitudes and colors from 3 consecutive nights after rotation correction.

| Date | B | V | R | I | B-V | V-R | R-I | V-I |
|---|---|---|---|---|---|---|---|---|
| 1/25/2005 | 1.137 ±0.036 | 0.445 ±0.042 | 0.143 ±0.028 | -0.183 ±0.023 | 0.692 ±0.055 | 0.302 ±0.050 | 0.326 ±0.037 | 0.628 ±0.048 |
| 1/25/2005 | 1.162 ±0.028 | 0.562 ±0.046 | 0.151 ±0.039 | -0.159 ±0.023 | 0.600 ±0.054 | 0.411 ±0.061 | 0.310 ±0.045 | 0.721 ±0.052 |
| 1/26/2005 | 1.087 ±0.027 | 0.505 ±0.047 | 0.111 ±0.037 | -0.170 ±0.024 | 0.582 ±0.055 | 0.394 ±0.060 | 0.281 ±0.044 | 0.675 ±0.053 |
| 1/26/2005 | 1.202 ±0.096 | 0.424 ±0.037 | 0.093 ±0.023 | -0.146 ±0.026 | 0.778 ±0.103 | 0.331 ±0.044 | 0.239 ±0.035 | 0.570 ±0.045 |
| 1/27/2005 | 1.145 ±0.027 | 0.542 ±0.032 | 0.136 ±0.029 | -0.247 ±0.031 | 0.603 ±0.042 | 0.406 ±0.043 | 0.383 ±0.042 | 0.789 ±0.044 |
| 1/27/2005 | | 0.449 ±0.037 | 0.193 ±0.023 | -0.264 ±0.028 | | 0.256 ±0.044 | 0.457 ±0.037 | 0.713 ±0.047 |
| **Weighted Average** | 1.133 ±0.014 | 0.487 ±0.016 | 0.139 ±0.012 | -0.188 ±0.010 | 0.626 ±0.025 | 0.343 ±0.020 | 0.333 ±0.016 | 0.683 ±0.020 |

Table 4. Solution for 2003 EL61's density, ρ, semi-major axes ($a_1$, $a_2$, and $a_3$), and visual albedo, $p_v$, for given values of $a_1/a_2$ and angle between the spin axis and line of site, φ.

| φ (°) | $a_1/a_2$ | ρ (kg m$^{-3}$) | $a_1$ (km) | $a_2$ (km) | $a_3$ (km) | $p_v$ |
|---|---|---|---|---|---|---|
| 74.4 | 1.316 | 2612 | 989 | 752 | 497 | 0.714 |
| 67.1 | 1.389 | 2646 | 1013 | 730 | 493 | 0.679 |
| 62.4 | 1.471 | 2690 | 1039 | 707 | 488 | 0.653 |
| 58.8 | 1.563 | 2744 | 1067 | 683 | 482 | 0.633 |
| 55.8 | 1.667 | 2812 | 1096 | 658 | 476 | 0.619 |
| 53.2 | 1.786 | 2896 | 1128 | 632 | 467 | 0.608 |
| 51.0 | 1.923 | 3002 | 1162 | 604 | 458 | 0.602 |
| 49.0 | 2.083 | 3134 | 1198 | 575 | 446 | 0.599 |
| 47.1 | 2.273 | 3302 | 1238 | 545 | 433 | 0.599 |
| 46.8 | 2.314 | 3340 | 1246 | 539 | 430 | 0.600 |
| 90.0 | 1.29 ± 0.04 | 2600 ± 37 | 980 ± 27 | 759 ± 24 | 498 ± 4 | 0.726 ± 0.01 |

Note: The values listed in the last row are extrapolated from the values in the preceding rows. Their errors derive from the uncertainty in $a_1/a_2$.

Table 5. Derived density (ρ), dimensions ($a_1$, $a_2$, and $a_3$), and visual albedo ($p_v$) for 2003 EL61 assuming different shape models.

| Shape model | ρ (kg m$^{-3}$) | dimensions (km) | length (km) | $p_v$ |
|---|---|---|---|---|
| **triaxial ellipsoid, $\phi = 90°$** | 2600 | $a_1$=980, $a_2$=759, $a_3$=498 | 1960 | 0.73 |
| **triaxial ellipsoid, $\phi > 47°$** | <3340 | $a_1$<1250, $a_2$>540, $a_3$>430 | < 2500 | 0.60 |
| **oblate spheroid** | >2530 | $a_1$ < 870, $a_3$ > 500 | <1640 | >0.71 |
| **oblate spheroid** | <3300 | $a_1$ > 750, $a_3$ < 524 | >1500 | <0.80 |

Notes: For the triaxial ellipsoid, $a_1$, $a_2$, and $a_3$ are the semi-major axes; for the oblate spheroid, $a_1$ and $a_3$ are the equatorial and polar radii.

Figure Captions

Fig. 1. Reduced magnitudes for 2003 EL61 versus Julian Date (uncorrected for rotational variation). Filled diamonds, filled squares, crosses, and filled triangles respectively represent SMARTS 1.3m observations in Johnson/Cousins B, V, R, and I. Open circles and open squares represent the Palomar 200" and Tenagra 32" data sets which are normalized so that their mean's are zero. See text for details.

Figure 2. Dispersion, θ, calculated by the phase-dispersion method as a function of rotational period. Periods yielding the lowest θ are the most likely rotation periods. Figure 3. Relative brightness of the combined data from Figure 1 versus rotational phase assuming period 0.163135 days. Small diamonds (blue), boxes (green), and triangles (red) represent the SMARTS B, V, and I data, respectively. Large circles and boxes represent the respective Palomar and Tenagra data sets. The dashed line shows a fit to the light curve obtained median averaging the data in 40 equally sized bins in rotational phase. The dependence on solar phase angle has been removed from the SMARTS data. See text for details.

Figure 3. Relative brightness of the combined data from Figure 1 versus rotational phase assuming period 3.9154 h. Small diamonds (blue), boxes (green), and triangles (red) represent the SMARTS B, V, and I data, respectively. Large circles and boxes represent the respective Palomar and Tenagra data sets. The dashed line shows a fit to the light curve obtained median averaging the data in 40 equally sized bins in rotational phase. The dependence on solar phase angle has been removed from the SMARTS data. See text for details.

Figure 4. Relative brightness in B, V, and I (diamonds, squares, and triangles, respectively) versus rotational phase. These data are derived from the SMARTS observations plotted in Fig 3 after binning by rotational phase and evaluating the median average within each bin. See text for details.

Figure 5. Average reduced magnitude versus solar phase angle for the B, V, and I band SMARTS observations of 2003 EL61 (diamonds, squares, and triangles, respectively). The error bars for each data point are smaller than the data symbols.

Figure 6. V-I minus the sun's value (V-I = 0.69) for the known KBOs with absolute magnitudes less than 4. Values for absolute magnitude are as listed by the Minor Planet Center (http://cfa-www.harvard.edu/iau/mpc.html ). Values for V-I are as reported by this paper for 2003 $EL_{61}$, by Stern & Yelle (1999) for Pluto and Charon, by Boehnhardt et al (2004) for Ixion , by Jewitt & Shepard (2002) for Varuna, by Brown, Trujillo & Rabinowitz (2005) for 2003 $UB_{313}$, and by Schaefer et al. (2005) for all others.

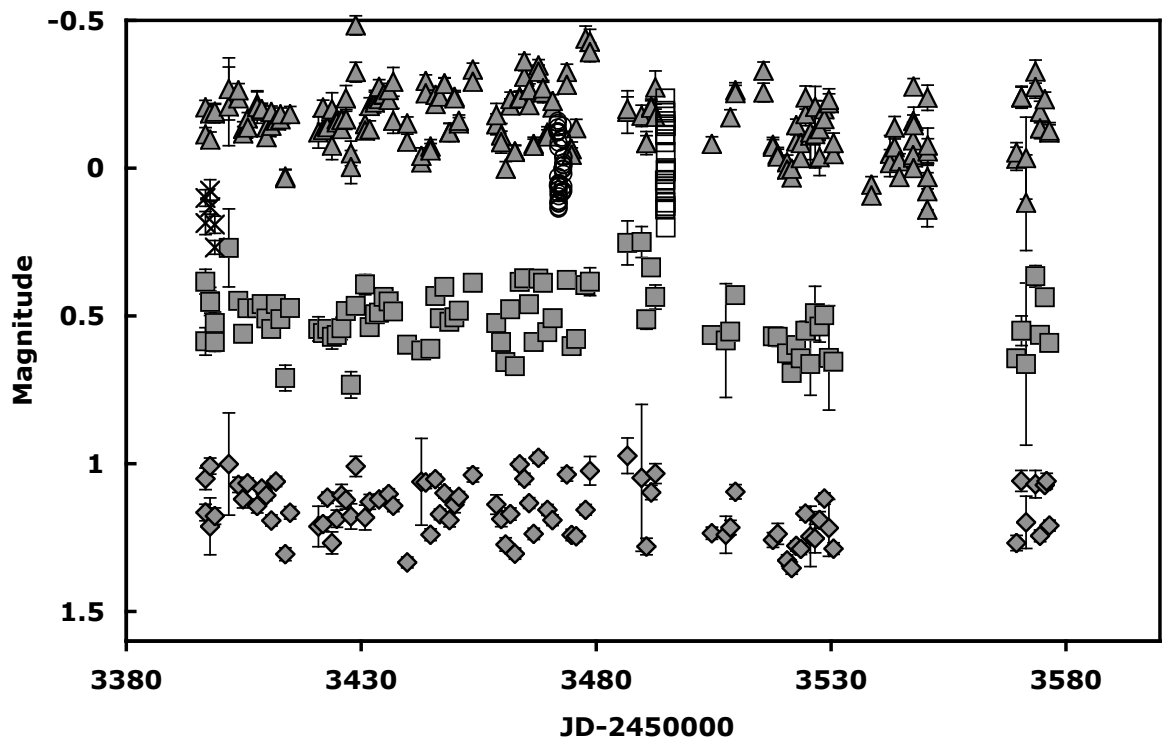

Fig. 1. Reduced magnitudes for 2003 EL61 versus Julian Date (uncorrected for rotational variation). Filled diamonds, filled squares, crosses, and filled triangles respectively represent SMARTS 1.3m observations in Johnson/Cousins B, V, R, and I. Open circles and open squares represent the Palomar 200" and Tenagra 32" data sets which are normalized so that their mean's are zero. See text for details.

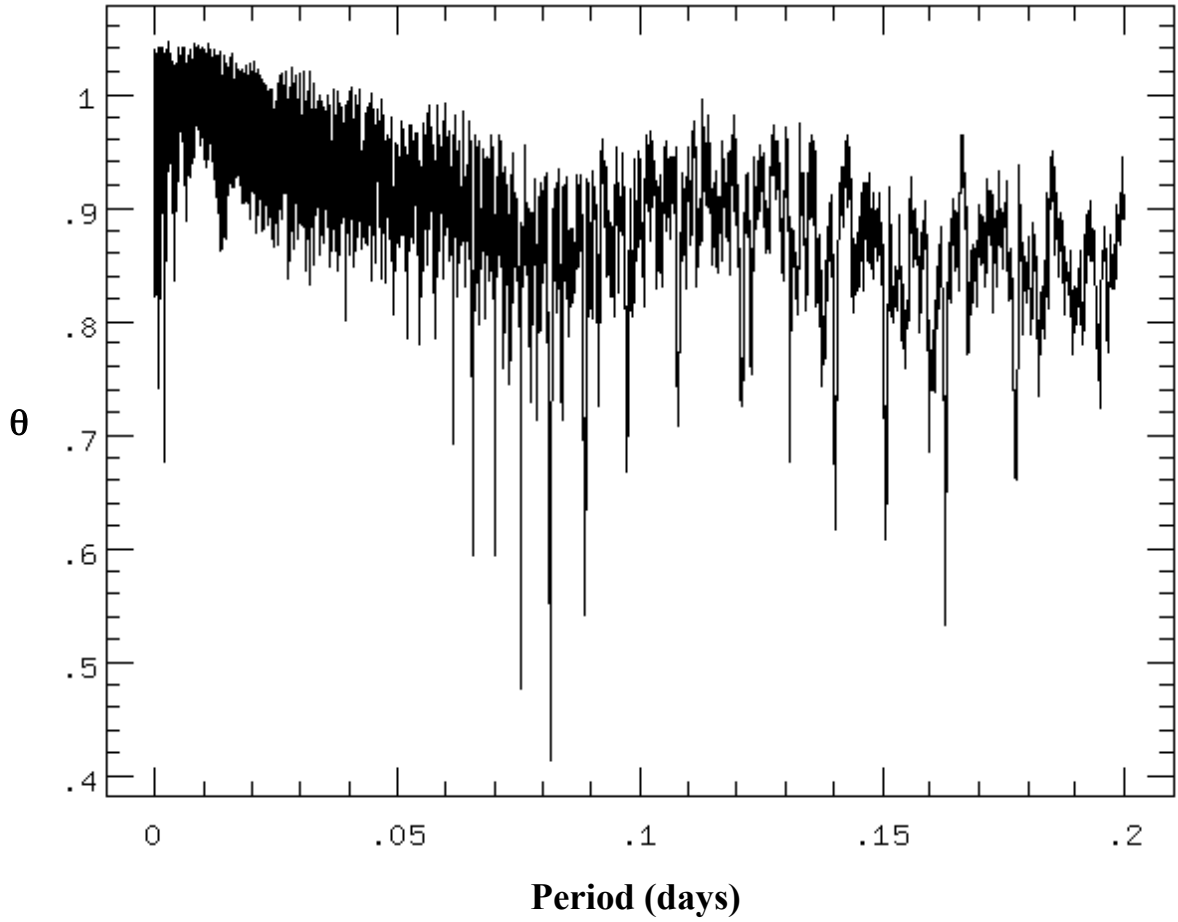

Figure 2. Dispersion, θ, calculated by the phase-dispersion method as a function of rotational period. Periods yielding the lowest θ are the most likely rotation periods.

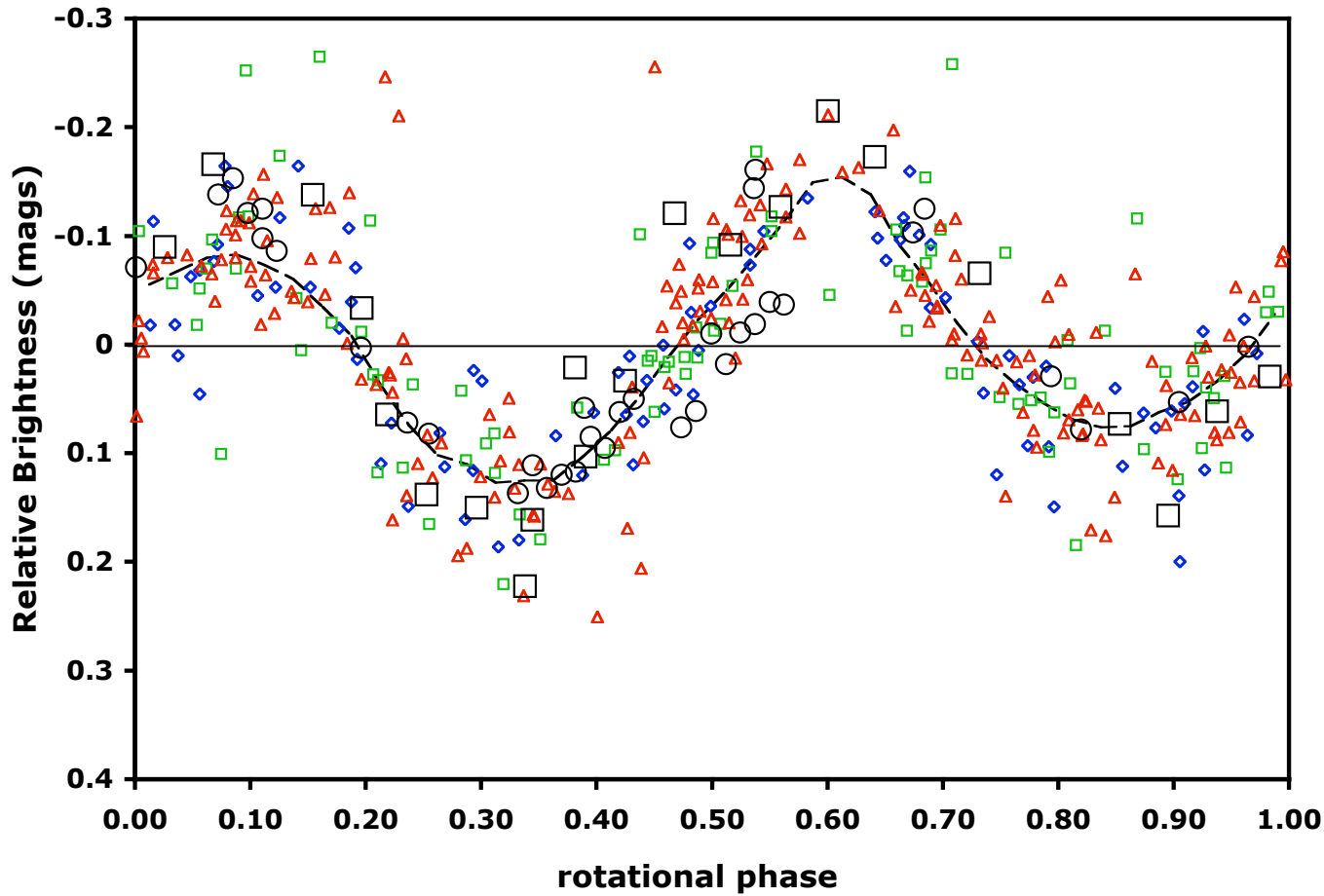

Figure 3. Relative brightness of the combined data from Figure 1 versus rotational phase assuming period 3.9154 h. Small diamonds (blue), boxes (green), and triangles (red) represent the SMARTS B, V, and I data, respectively. Large circles and boxes represent the respective Palomar and Tenagra data sets. The dashed line shows a fit to the light curve obtained median averaging the data in 40 equally sized bins in rotational phase. The dependence on solar phase angle has been removed from the SMARTS data. See text for details.

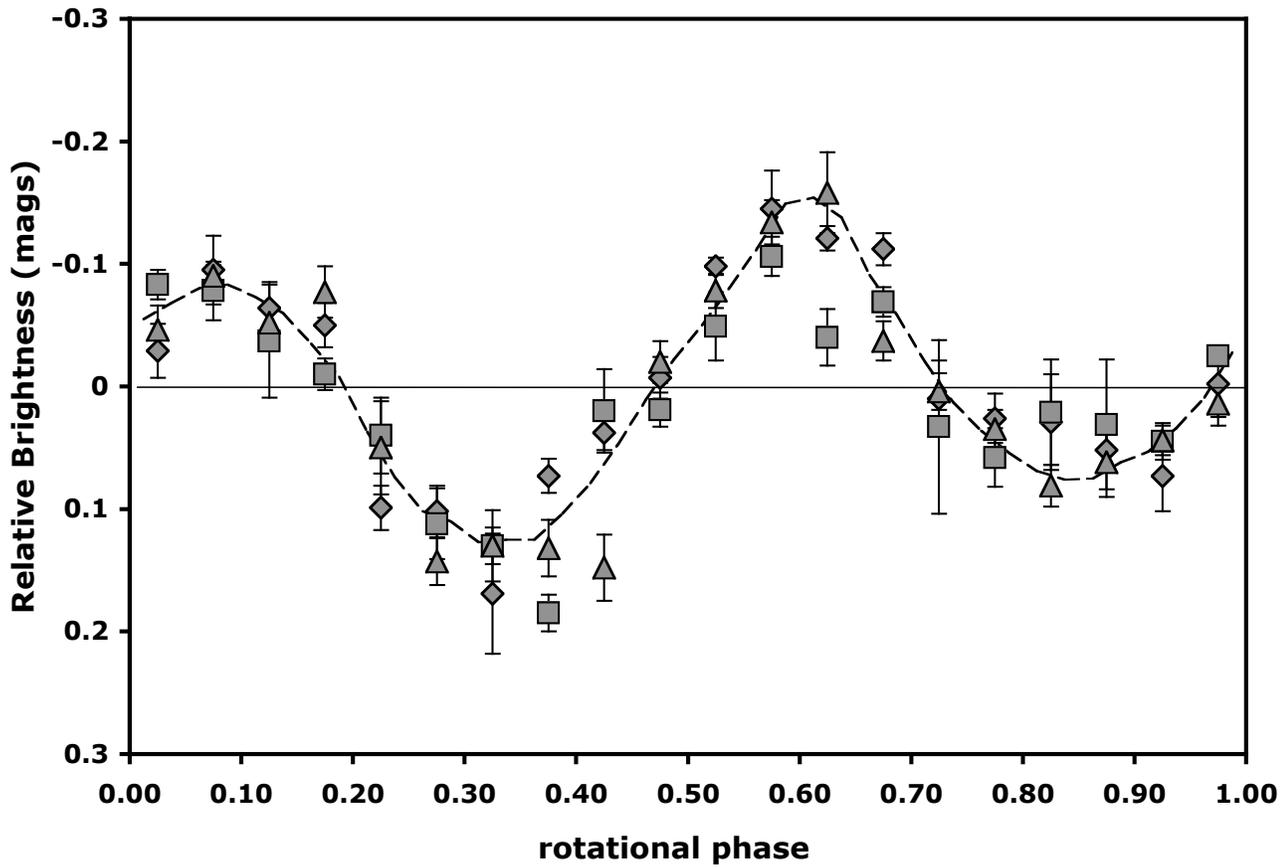

Figure 4. Relative brightness in B, V, and I (diamonds, squares, and triangles, respectively) versus rotational phase. These data are derived from the SMARTS observations plotted in Fig 3 after binning by rotational phase and evaluating the median average within each bin. See text for details.

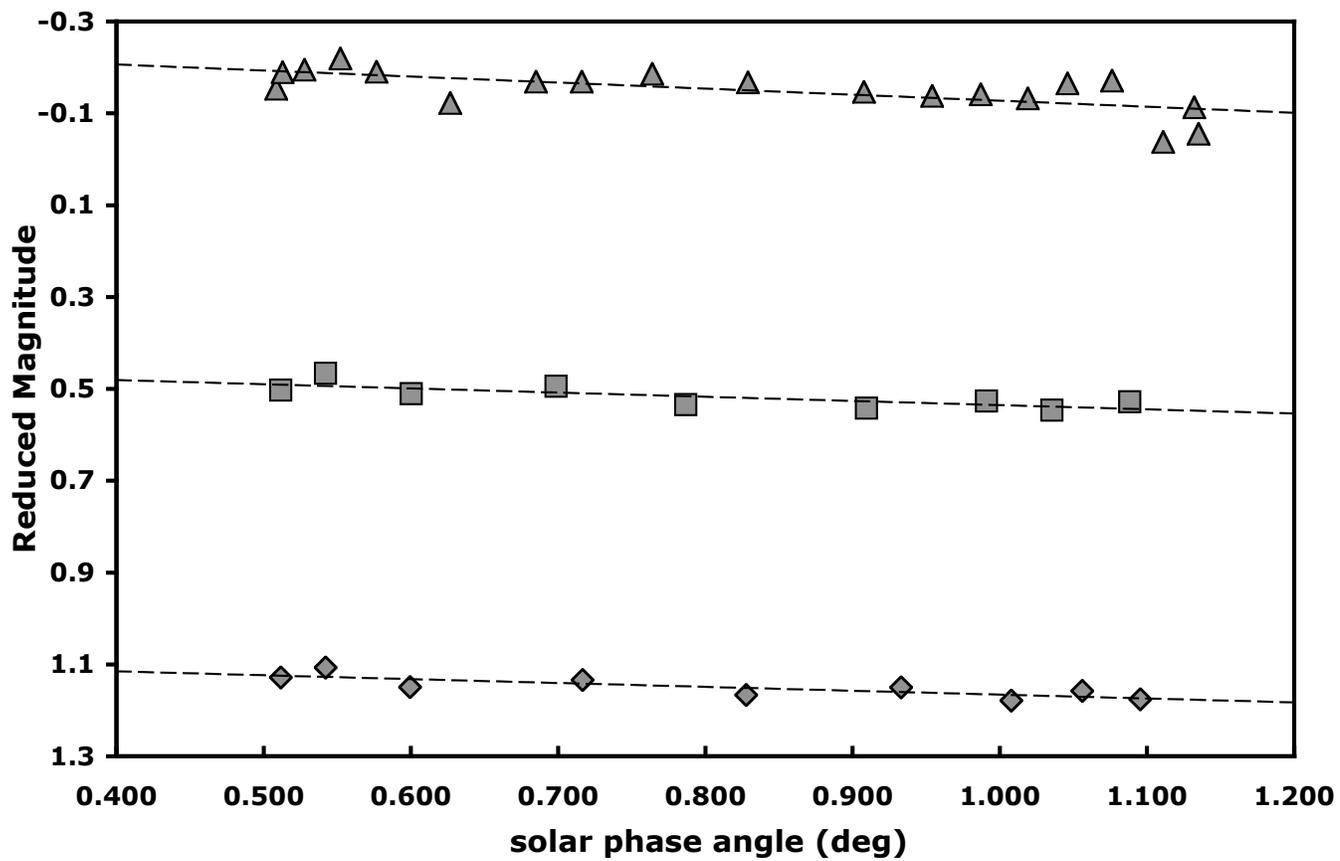

Figure 5. Average reduced magnitude versus solar phase angle for the B, V, and I band SMARTS observations of 2003 EL61 (diamonds, squares, and triangles, respectively). The error bars for each data point are smaller than the data symbols.

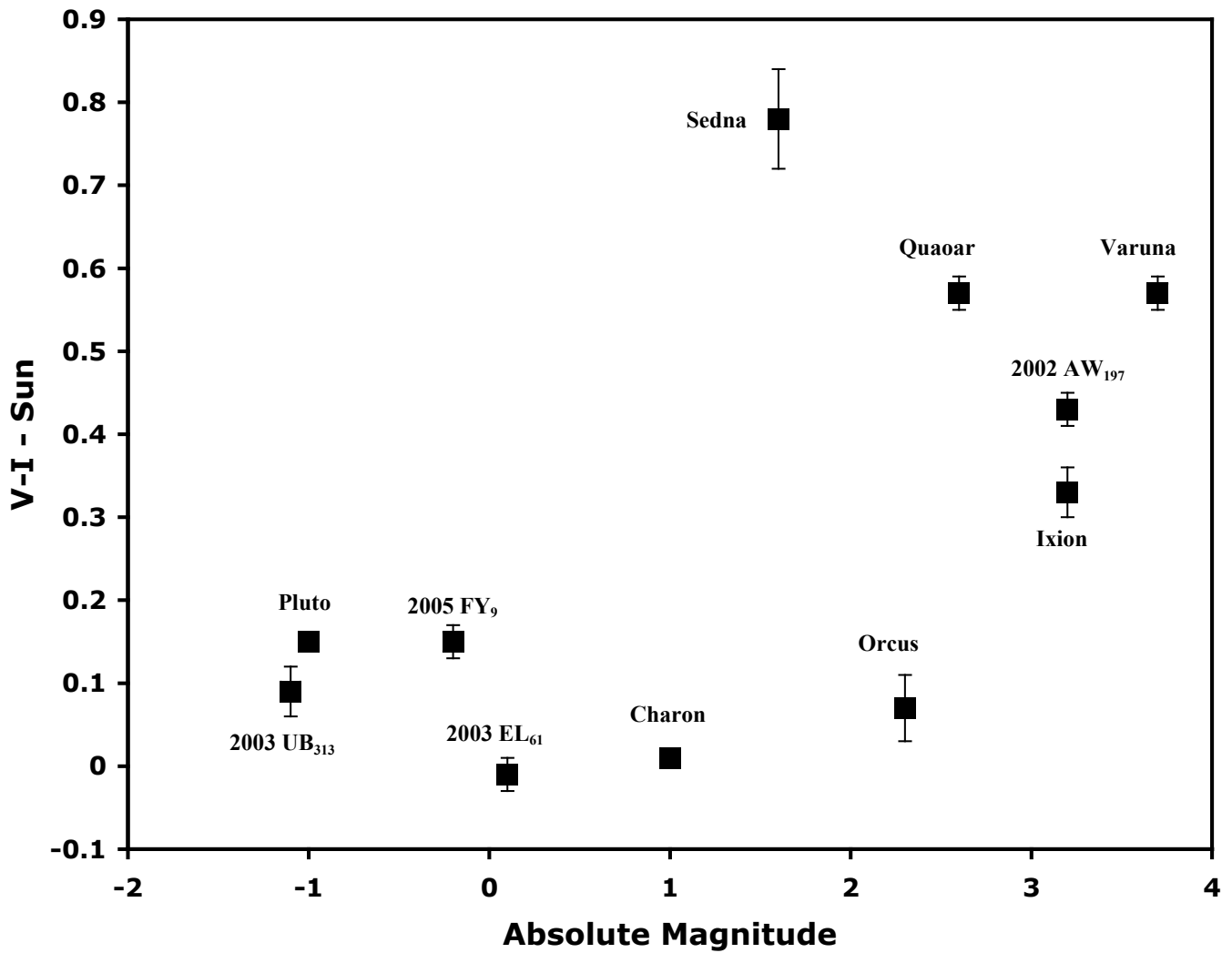

Figure 6. V-I minus the sun's value (V-I = 0.69) for the known KBOs with absolute magnitudes less than 4. Values for absolute magnitude are as listed by the Minor Planet Center (http://cfa-www.harvard.edu/iau/mpc.html ). Values for V-I are as reported by this paper for 2003 $EL_{61}$, by Stern & Yelle (1999) for Pluto and Charon, by Boehnhardt et al (2004) for Ixion , by Jewitt & Shepard (2002) for Varuna, by Brown, Trujillo & Rabinowitz (2005) for 2003 $UB_{313}$, and by Schaefer et al. (2005) for all others.